\documentclass[lettersize,journal]{IEEEtran}
\usepackage{amsmath,amsfonts}
\usepackage{algorithmic}
\usepackage{algorithm}
\usepackage{array}
\usepackage[caption=false,font=normalsize,labelfont=sf,textfont=sf]{subfig}
\usepackage{textcomp}
\usepackage{stfloats}
\usepackage{url}
\usepackage{verbatim}
\usepackage{graphicx}
\usepackage{cite}
\usepackage{xcolor}
\usepackage{booktabs} 
\usepackage{multirow}
\usepackage[table]{xcolor}
\usepackage{pifont}

\usepackage{hyperref}

\begin{document}

\title{Perception-Consistency Multimodal Large Language Models Reasoning via Caption-Regularized Policy Optimization}

\author{Songjun Tu$^{1,2,3}$,
        Qichao Zhang$^{1,3}$,
        Jingbo Sun$^{1,2,3}$,
        Yuqian Fu$^{1,3}$,
        Linjing Li$^{1,3}$, 
        Xiangyuan Lan$^{2}$, \\ 
        Dongmei Jiang$^{2}$, 
        Yaowei Wang$^{4,2}$ and 
        Dongbin Zhao$^{1,2,3}$,~\IEEEmembership{Fellow,~IEEE}%
        
\thanks{$1$ State Key Laboratory of Multimodal Artificial Intelligence Systems, Institute of Automation, Chinese Academy of Sciences, Beijing, 100190, China. (tusongjun2023@ia.ac.cn, zhangqichao2014@ia.ac.cn, sunjingbo2022@ia.ac.cn, fuyuqian2022@ia.ac.cn, linjing.li@ia.ac.cn, dongbin.zhao@ia.ac.cn)}%

\thanks{$2$ Pengcheng Laboratory, Shenzhen, 518055, China. (lanxy@pcl.ac.cn, jiangdm@pcl.ac.cn)}%

\thanks{$3$ School of Artificial Intelligence, University of Chinese Academy of Sciences, Beijing, 100049, China. }%

\thanks{$4$ Harbin Institute of Technology (Shenzhen), Shenzhen, 518055, China. (wangyaowei@hit.edu.cn)}%


\thanks{All authors declare that they have no conflict of interest.}%

}



\maketitle

\begin{abstract}
While multimodal large language models excel at tasks that integrate visual perception with symbolic reasoning, their performance is often undermined by a critical vulnerability: perception-induced errors that propagate through the reasoning chain. 
Current reinforcement learning (RL) fine-tuning methods, while enhancing reasoning abilities, largely fail to address the underlying misalignment between visual grounding and the subsequent reasoning process.
To address this challenge, we propose \textbf{Caption-Regularized Policy Optimization (CapPO)}, a novel RL framework that explicitly enforces perceptual consistency during policy optimization. 
CapPO integrates two key mechanisms: 
(1) a caption-based consistency regularization, which minimizes the divergence between responses conditioned on raw images and those conditioned on captions, thereby anchoring reasoning to semantically faithful visual content; and 
(2) a KL-weighted advantage estimation scheme, which adaptively scales reinforcement signals to strengthen perceptually consistent trajectories while suppressing spurious correlations. 
Extensive experiments on five math-focused and five general reasoning benchmarks demonstrate that CapPO achieves competitive performance, yielding gains of +6.0\% accuracy on math-related tasks and +2.4\% on general reasoning tasks over the base Qwen2.5-VL-7B model. 
Moreover, ablation studies further confirm the effectiveness of each component, while error analysis reveals that CapPO significantly reduces perception-related mistakes compared with baselines. 
Overall, CapPO provides a simple yet effective framework for improving multimodal reasoning.
\end{abstract}


\section{Introduction}
Multimodal large language models (MLLMs) have become key enablers for tasks requiring both visual perception and symbolic reasoning, showing strong potential in domains such as autonomous driving \cite{liu2025reasonplan,zheng2024planagent}, robotics~\cite{li2024manipllm,chen2025robogpt}, and 3D scene understanding~\cite{liu2023multi, xiong20253ur}, where reliable reasoning demands accurate perception working seamlessly with dependable reasoning, ensuring consistent and trustworthy outcomes.

Recent progress in reinforcement learning (RL) has demonstrated that verifiable rewards can effectively shape long-chain reasoning behaviors in LLMs and MLLMs~\cite{guo2025deepseek,tu2025learning,fu2025rlae,xie2025logic}.
Based on this idea, recent methods further incorporate iterative refinement~\cite{deng2025openvlthinker}, rule-based verification~\cite{meng2025mm}, and reward shaping~\cite{wang2025vl}, which collectively enhance reasoning stability, output correctness, and training efficiency.
Nevertheless, perception errors in vision encoders remain a major bottleneck, as inaccuracies in visual grounding can cascade into hallucinations or logically inconsistent outputs~\cite{liu2024survey}.


\begin{figure}[t]
\centering
\includegraphics[width=\columnwidth]{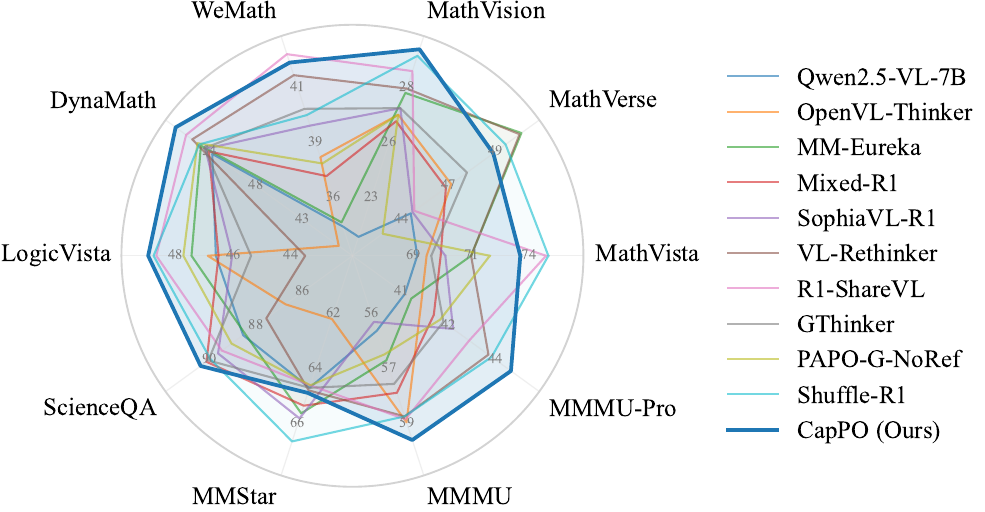}
\caption{Accuracy comparison of CapPO with baseline models on math-related and general reasoning benchmarks.}
\label{fig_ladar}
\vspace{-1em}
\end{figure}

Reasoning-centric RL methods improve textual correctness and consistency but largely assume accurate perceptual inputs, whereas perception-focused approaches enhance grounding without explicitly linking it to downstream reasoning~\cite{zhou2025reinforced, wang2025perception}.
Some recent works, such as PAPO~\cite{wang2025perception} and Vision Matters~\cite{li2025vision}, attempt to bridge this gap by introducing visual perturbations and enforcing divergence between answers conditioned on original versus perturbed images, thereby encouraging MLLMs to ground responses in actual visual content.
While these strategies yield certain gains, models may still exploit spurious correlations, producing reasoning traces that appear coherent yet rest on flawed perceptual grounding.
Addressing this misalignment between perception and reasoning remains a central obstacle to advancing MLLM reliability.
Benchmark studies such as \textit{MathVista}~\cite{lu2023mathvista} and \textit{MMMU}~\cite{yue2024mmmu} further reveal that perception-heavy questions remain a dominant source of error, even when reasoning chains look logically sound.
Together, these observations underscore the need to explicitly incorporate perceptual consistency into the learning objective rather than treating perception and reasoning as separate modules.

To bridge this gap, we propose \textbf{Cap}tion-Regularized \textbf{P}olicy \textbf{O}ptimization (\textbf{CapPO}), a RL framework that enforces perceptual consistency within policy optimization. 
The central idea is to use captions derived from the input image as lightweight semantic anchors, encouraging the model to ground its reasoning in reliable visual content rather than spurious correlations. 
By coupling caption-guided alignment with adaptive weighting, our method effectively suppresses perception-induced errors and achieves improvements across diverse evaluation settings.

Unlike prior methods that require heavy visual augmentation or adversarial perturbations, CapPO leverages captions that are readily available from MLLMs, making it both efficient and broadly applicable. 
Furthermore, the proposed KL-weighted advantage scheme adaptively balances correctness with perceptual alignment, giving stronger updates to trajectories grounded in reliable visual semantics.
This design enables CapPO to be seamlessly integrated into current RL pipelines while directly addressing the perception and reasoning gap.

In summary, our core contributions are as follows:

\begin{itemize}
    \item Our analysis indicates that perceptual errors constitute a primary source of reasoning failures in MLLMs, and that access to high quality captions alone can often yield superior answer accuracy (Section~\ref{sec:motivation}).
    \item We introduce a KL-based regularization that explicitly aligns responses conditioned on images with those conditioned on captions, thereby enforcing perceptual consistency (Section~\ref{sec:method1}).
    \item We propose an advantage-weighting scheme that adaptively modulates reinforcement signals based on perceptual consistency, giving stronger updates to trajectories grounded in reliable visual semantics (Section~\ref{sec:method2}).
    \item We achieve competitive performance on both math-related and general reasoning benchmarks (Fig.~\ref{fig_ladar}), with accuracy improvements of +6.0 and +2.4 over the base model, respectively (Section~\ref{sec:experiments}).
\end{itemize}

\section{Related Works}

\subsection{Reinforcement Learning for Multimodal Reasoning}
Reinforcement learning with verifiable rewards (RLVR) has emerged as a central paradigm for enabling long-chain reasoning in large models, with DeepSeek-R1 demonstrating that rule-based rewards can induce explicit reasoning behaviors~\cite{guo2025deepseek}. Reinforcement signals have also been shown to be effective in broader multimodal settings~\cite{zhou2025reinforced}.

Based on this idea, a number of approaches extend Group Relative Policy Optimization (GRPO) to multimodal large language models. Representative examples such as OpenVL-Thinker~\cite{deng2025openvlthinker} and VL-Rethinker~\cite{wang2025vl} incorporate iterative refinement and self-reflection mechanisms to enhance visual reasoning. Other variants adjust reward formulations or baseline sharing strategies, including Mixed-R1~\cite{xu2025mixed}, SophiaVL-R1~\cite{fan2025sophiavl}, R1-ShareVL~\cite{yao2025r1}, and GThinker~\cite{zhan2025gthinker}. Meanwhile, Vision-R1~\cite{huang2025vision} investigates cold-start reinforcement for multimodal reasoning. Collectively, these studies confirm that reinforcement signals are effective for coupling perception and reasoning in multimodal contexts.

More recent work has shifted toward reward shaping and efficiency. For instance, StepGRPO~\cite{zhang2025r1} introduces step-wise dense rewards to mitigate sparsity, while efficiency-oriented designs such as Shuffle-R1~\cite{zhu2025shuffle} improve rollout scalability and reduce supervision costs.

Distinct from these directions, our CapPO framework explicitly enforces perceptual consistency within the RL stage by aligning policies conditioned on images and captions, and by weighting advantages with KL divergence. This design suppresses trajectories that exploit perceptual errors and strengthens reliable reasoning.

\subsection{Perceptual Alignment and Consistency}
Accurate multimodal reasoning hinges on reliable perception and cross-modal alignment. Early studies showed that task-adaptive attention mechanisms can improve image captioning by focusing on semantically relevant regions, thereby strengthening visual grounding~\cite{liu2022depth, ma2023boosting}. Subsequent advances further emphasized perceptual alignment, where curriculum-based adaptation methods were shown to suppress irrelevant signals and enhance consistency between vision and language representations~\cite{xiao2023clip}.

Despite these advances, most methods strengthen perception or cross-modal alignment in isolation rather than enforcing consistency within RL objectives. Large-scale contrastive pretraining has compacted vision–language representations and improved downstream recognition and retrieval~\cite{radford2021learning,jia2021scaling}, while modern multimodal encoders leverage caption-based pretraining and self-distillation to enhance aligned features~\cite{tschannen2025siglip}. 
These approaches support the premise that better perception leads to better reasoning, yet they stop short of imposing perceptual consistency during policy optimization.
To address this gap, recent perception-aware approaches~\cite{wang2025perception,li2025vision} highlight the role of accurate and stable perception but still fall short of fully resolving the misalignment between perception and reasoning.

A newer strand explicitly couples perception with reasoning, such as grounded chains of thought that tie rationales to image regions to mitigate hallucination~\cite{fan2025grit,wu2025grounded} and analyses of fast–slow reasoning that link reasoning dynamics with perceptual signals~\cite{xiao2025fast}.
Yet these methods typically depend on heavy supervision or detectors, whereas consistency-aware RL has only recently emerged as a more direct means of optimizing perceptual consistency within the learning loop~\cite{chen2025grpo,yu2025perception}.

In contrast, our CapPO uses captions as lightweight anchors to implicitly enforce perception alignment inside the RL objective, avoiding external detectors and costly labels while directly regularizing policy learning.

\section{Preliminaries}

\subsection{Group Relative Policy Optimization}
Group Relative Policy Optimization (GRPO) is an efficient and effective RL algorithm for LLM post-training, especially in verifiable tasks.
Let $q \in \mathcal{Q}$ denote a question (input prompt), and let $\pi_\theta(o \mid q)$ be a policy parameterized by $\theta$. For each $q$, GRPO samples a group of $G$ candidate outputs $\{o_1, o_2, \dots, o_G\} \sim \pi_{\theta_{\text{old}}}(\cdot \mid q)$ from the old policy $\pi_{\theta_{\text{old}}}$. A reward model or rule-based verifier $r_\phi$ then assigns each output $o_i$ a token-level reward $r_i$.

Then, GRPO computes a \textit{group-relative advantage} by normalizing rewards within the sampled group as 
\begin{equation}
\hat{A}_i = (r_i - \mu)/\sigma
\label{eq:a}
\end{equation}
where $\mu = \tfrac{1}{G}\sum_{j=1}^G r_j$ and $\sigma = \sqrt{\tfrac{1}{G}\sum_{j=1}^G (r_j - \mu)^2}$. 
This group-based normalization removes the need for a separate critic network, thereby simplifying training and reducing memory cost, while ensuring that outputs with higher-than-average rewards receive positive advantages and those with lower rewards are penalized.

The optimization objective of GRPO is defined as
\begin{equation}
\label{eq:grpo}
\begin{aligned}
J&_{\mathrm{GRPO}}(\theta) = \;  
\mathbb{E}_{q, \{o_i\}_{i=1}^G} \Bigg[
\frac{1}{G} \sum_{i=1}^G \frac{1}{|o_i|} \sum_{t=1}^{|o_i|} \Big( \min\!\big( \rho_{i,t}\hat{A}_{i}, \; \\
& \text{clip}(\rho_{i,t}, 1-\epsilon, 1+\epsilon)\hat{A}_{i}\big) 
- \gamma D_{\mathrm{KL}}(\pi_\theta \parallel \pi_{\mathrm{ref}}) \Big) \Bigg],
\end{aligned}
\end{equation}
where $\rho_{i,t} = \tfrac{\pi_\theta(o_{i,t}\mid q, o_{i,<t})}{\pi_{\theta_{\text{old}}}(o_{i,t}\mid q, o_{i,<t})}$ is the importance sampling ratio, $\epsilon$ is the clipping threshold, $D_{\mathrm{KL}}$ denotes the Kullback–Leibler (KL) divergence, $\gamma$ is the KL coefficient, and the reference policy $\pi_{\mathrm{ref}}$ is typically the supervised fine-tuned (SFT) model. 

By employing a group-based baseline for advantage estimation, GRPO enables stable and efficient optimization without learning an explicit value function, making it well-suited for tasks with verifiable rewards such as mathematics or logic reasoning. 
In this work, we adopt GRPO as the fundamental training paradigm and introduce additional modules to explicitly enforce perceptual consistency.

\begin{figure}[t]
\centering
\includegraphics[width=\linewidth]{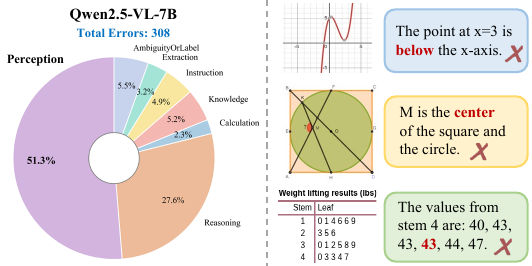}
\caption{(Left) Error analysis of Qwen2.5-VL-7B on MathVista. 
Over half of the errors (51.3\%) stem from perceptual failures. 
(Right) Examples on the right illustrate typical cases such as curve misjudgment, geometric misidentification, and misreading of stem and leaf plots.
}
\vspace{-1em}
\label{fig_qwen_error}
\end{figure}

\subsection{Problem Setup}
We consider a multimodal reasoning task defined over paired image–question inputs. In addition to the natural language question $q \in \mathcal{Q}$, the input also contains one or multiple images $i \in \mathcal{I}$. 
The policy model $\pi_\theta(o \mid i, q)$ parameterized by $\theta$ generates a response $o$ conditioned on both the image(s) and the question. 
The primary objective of this work is to improve the reliability and accuracy of MLLMs in answering questions based on the provided images. 
To incorporate semantic grounding from visual content, we further introduce a captioner $C(\cdot)$ that produces a textual description $c = C(i,q)$ for each image. This yields an alternative conditional distribution $\pi_\theta(o \mid c, q)$, where the model generates the response conditioned on the caption–question pair instead of the raw image–question pair.

\section{Key Observations}
\label{sec:motivation}

\subsection{Perceptual Errors as the Fundamental Bottleneck}
We first conduct an error analysis on \textit{Qwen2.5-VL-7B} \cite{bai2025qwen2} using the \textit{MathVista} \cite{lu2023mathvista} dataset. 
The results reveal that perceptual failures are the dominant source of errors, 
with 51.3\% of mispredictions arising from incorrect visual grounding, as shown in Fig.~\ref{fig_qwen_error}. 
For illustration, the figure also presents representative excerpts from the model’s reasoning traces, 
where the errors clearly originate from flawed perception, such as misreading diagrams, misidentifying geometric relations, or overlooking key attributes in figures and tables. 
These perceptual errors are particularly detrimental because they propagate through the reasoning chain: 
once the model builds its logic on flawed perception, the subsequent steps may appear consistent yet inevitably lead to incorrect answers. 
This observation indicates that improving reasoning alone is insufficient; perceptual grounding is essential for ensuring reliable multimodal reasoning.

\subsection{Captions as Effective Substitutes for Images in Reasoning}

\begin{table}[!t]
\centering
\caption{Accuracy of DeepSeek-R1-0528-Qwen3-8B on \textit{ViRL39K} using captions generated by Qwen2.5-VL of different scales.}
\label{tab:caption_results}
\resizebox{.9\columnwidth}{!}{
\begin{tabular}{lcccc}
\toprule
\textbf{Caption Model Size} & \textbf{3B} & \textbf{7B} & \textbf{32B} & \textbf{72B} \\
\midrule
Reasoning Accuracy (\%) & 66.1 & 70.4 & 75.1 & 75.3 \\
\bottomrule
\end{tabular}
}
\vspace{-1em}
\end{table}

To further investigate the role of perceptual grounding in multimodal reasoning, 
we examine whether high-quality captions can serve as effective substitutes for raw images in downstream reasoning tasks. 
Specifically, we employ \textit{Qwen2.5-VL} models of different scales (3B, 7B, 32B, and 72B) to generate captions on the \textit{ViRL39K} dataset \cite{wang2025vl}. 
These captions, together with the original questions, are then provided as input to a purely text-based reasoning model, 
\textit{DeepSeek-R1-0528-Qwen3-8B} \cite{guo2025deepseek}. 
In this setting, reasoning errors can no longer be attributed to visual perception within the reasoning trace, 
as perception is delegated entirely to the caption generation step. 
Table~\ref{tab:caption_results} shows that captions generated by stronger models (32B, 72B) allow the text-only reasoning model to achieve over 75\% accuracy, demonstrating that high-quality captions can provide sufficient semantic grounding for reasoning. 

Nevertheless, directly replacing images with captions during multimodal training is problematic: 
text supervision cannot update the vision encoder, and bypassing the image entirely fails to strengthen the model’s perceptual grounding. 
Therefore, we argue that \textbf{\textit{caption information should be leveraged as a complementary signal during multimodal training to enhance perception}}, rather than as a wholesale substitute for the visual input.

\section{Method}

In this section, we term our framework \textbf{CapPO} (\textbf{Cap}tion-Regularized \textbf{P}olicy \textbf{O}ptimization), which integrates group-based policy optimization with caption-guided regularization for multimodal reasoning tasks. 

\begin{figure*}[t]
\centering
\includegraphics[width=\textwidth]{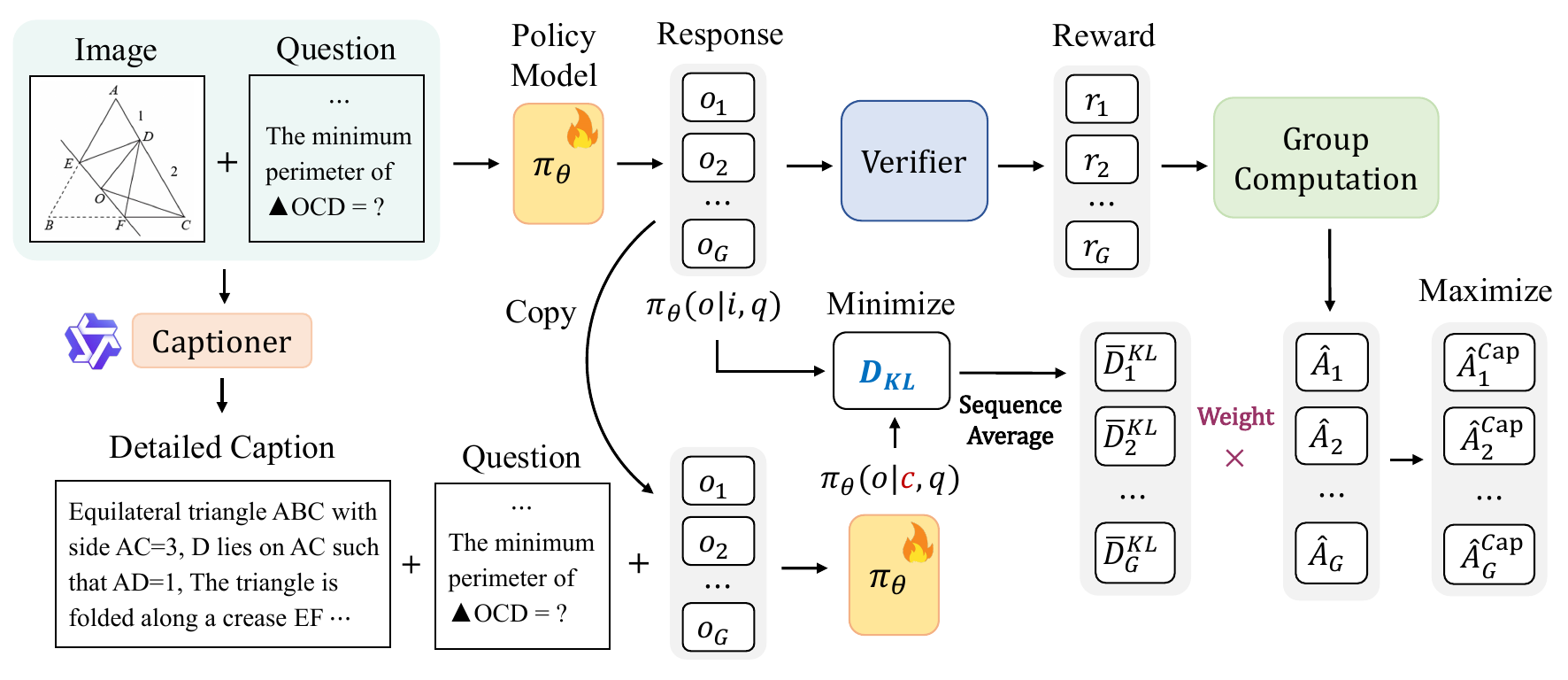}
\caption{Overview of the proposed Caption-Regularized Policy Optimization (CapPO) framework. CapPO extends GRPO by introducing caption-based consistency regularization (Section \ref{sec:method1}) and KL-weighted advantage estimation (Section \ref{sec:method2}), aligning policies conditioned on images and captions to suppress perceptual errors while enhancing reasoning accuracy.}
\label{fig_main}
\vspace{-1em}
\end{figure*}

\subsection{Overview}
An overview of the framework is shown in Fig.~\ref{fig_main}. 
Given an input consisting of one or multiple images $i \in \mathcal{I}$ and an associated question $q \in \mathcal{Q}$, the policy model $\pi_\theta$ generates candidate responses $o$. 
Depending on whether the image is used directly or replaced with its textual caption $c = C(i,q)$, we obtain two conditional distributions. 
The direct image–question policy is optimized with group-based reinforcement signals (Section \ref{sec:reward}), while the caption-conditioned policy serves as a semantic reference to regularize the learning process (Section \ref{sec:method1}). 
In addition, KL-weighted advantage estimation (Section \ref{sec:method2}) adaptively scales the reinforcement signals to emphasize perceptually consistent trajectories. Finally, these components are combined into the overall CapPO objective (Section \ref{sec:overall}).


\subsection{Reward Function}
\label{sec:reward}
Following prior works \cite{wang2025vl,wang2025perception}, we use a simple reward function to supervise both reasoning format and final correctness. 
For each input question, the model is required to generate a reasoning trace wrapped in \texttt{<think></think>} and extract the final answer within \texttt{\textbackslash boxed\{\}}. 
The reward for a question $q$ and a response $o$ is defined as
\begin{equation}
\label{eq:reward}
r(o \mid q) =
\begin{cases}
1.0, & \text{if $o$ is well-formatted and correct}, \\
0.1, & \text{if $o$ is well-formatted but wrong}, \\
0,   & \text{otherwise},
\end{cases}
\end{equation}
where correctness is automatically judged by a rule-based verifier \cite{guo2025deepseek}, as all questions in our dataset have verifiable final answers.
This scheme encourages both structural compliance and correctness.

\subsection{Caption-based Consistency Regularization}
\label{sec:method1}
A core feature that differentiates CapPO from GRPO is the introduction of \textit{caption-based consistency regularization}. Specifically, given an image $i$ and question $q$, we construct two distributions: $\pi_\theta(o \mid i, q)$ and $\pi_\theta(o \mid c, q)$, where $c = C(i,q)$ denotes a caption produced by an external captioner. 

The captioner is a large MLLM employed during the offline construction of the RL dataset. 
It takes both the image $i$ and the question $q$ as inputs, together with the manually designed prompts, to generate a detailed textual description of the image. 
The resulting caption $c$ is purely text-based and provides sufficient visual information for answering the associated question. 
Importantly, the caption is not used during inference but only serves as auxiliary supervision during training, thus avoiding any additional inference overhead.

Since the caption $c$ provides reliable perceptual information, reasoning conditioned on $(c, q)$ reduces the risk of incorrect visual perception. 
Ideally, a model with accurate perception should exhibit similar policy distributions when answering a question based on the raw image $(i, q)$ and on the corresponding textual description $(c, q)$. 
To enforce this perceptual consistency, we minimize the KL divergence between the token-level distributions obtained from the two input conditions:
\begin{equation}
\label{eq:kl-loss}
\mathcal{L}_{\mathrm{cap}} = \mathbb{E}_{q, \{o_i\}_{i=1}^G} \left[ D_{\mathrm{KL}}\!\left(\pi_\theta(o_i \mid i, q) \;\|\; \pi_\theta(o_i \mid {\textcolor{blue}{c}}, q)\right) \right].
\end{equation}

This formulation measures the discrepancy in token-level probabilities between the two perceptual paths. 
In practice, following [1], we adopt an unbiased estimator of the KL divergence, which is always non-negative, defined as
\begin{equation}
\begin{aligned}
D_{\mathrm{KL}}&\left(\pi_\theta(o_i \mid i, q)  \parallel \pi_\theta(o_i \mid {\textcolor{blue}{c}}, q)\right)  =
\frac{\pi_\theta(o_{i,t} \mid {\textcolor{blue}{c}}, q, o_{i,<t})}{\pi_\theta(o_{i,t} \mid i, q, o_{i,<t})} \\
& - \log \frac{\pi_\theta(o_{i,t} \mid {\textcolor{blue}{c}}, q, o_{i,<t})}{\pi_\theta(o_{i,t} \mid i, q, o_{i,<t})} - 1.
\end{aligned}
\end{equation}

The loss term $\mathcal{L}_{\mathrm{cap}}$ is then incorporated into the GRPO objective (\ref{eq:grpo}) as an implicit perceptual regularizer, serving as a perceptual constraint unique to CapPO. 

\subsection{KL-weighted Advantage Estimation}
\label{sec:method2}

Beyond the consistency loss, we introduce \textit{KL-weighted advantage estimation} to further inject perceptual alignment into RL signals. 
In the GRPO setting, a rule-based verifier provides only a final reward at the last token of each sequence, while rewards at all other tokens are zero. 
Let $\hat{A}_{i}$ denote the standard group-relative advantage for output $o_i$ Equation (\ref{eq:a}). 
CapPO reweights advantages at the sequence level, so that sequences with smaller image–caption KL are assigned larger optimization weights. 

Let $M_i$ denote the set of response tokens. We first aggregate the token-level KL values over the response region to obtain a sequence-level perceptual KL:
\begin{equation}
\label{eq:seq-kl}
\bar{D}^{\mathrm{KL}}_{i} = \frac{1}{|M_i|}\sum_{t\in M_i} D^{\mathrm{KL}}_{i,t}.
\end{equation}

To make the sequence-level KL values comparable across different samples within the same group, we further apply group-wise normalization over $\{\bar{D}^{\mathrm{KL}}_{1}, \dots, \bar{D}^{\mathrm{KL}}_{G}\}$, using z-score normalization. This ensures that the weighting mechanism is calibrated locally within each GRPO group, rather than across the entire dataset.

Based on this quantity, we define a KL-adaptive weight that dynamically adjusts the contribution of each sequence according to its perceptual consistency. Specifically, sequences with large KL are considered perceptually inconsistent, and their optimization weight should be reduced if the advantage is positive, while the penalty should be amplified if the advantage is negative. This leads to the following sign-aware formulation:
\begin{equation}
\label{eq:weight}
w_{i} = \mathrm{clip}\!\left(\exp\!\big(-\beta \,\bar{D}^{\mathrm{KL}}_{i} \cdot \mathrm{sgn}(\hat{A}_{i})\big),\; w_{\min},\,w_{\max}\right),
\end{equation}
where $\beta>0$ controls the modulation strength and $0<w_{\min}\le 1 \le w_{\max}$ bound the weight within a stable range. 

Finally, the KL-weighted advantages are defined as
\begin{equation}
\label{eq:weighted-adv}
\hat{A}^{\mathrm{cap}}_{i} = w_{i}\cdot \hat{A}_{i}.
\end{equation}

Intuitively, when the image–caption discrepancy is large, $w_i$ decreases for $\hat{A}_{i}>0$ to avoid reinforcing inconsistent perception, and increases for $\hat{A}_{i}<0$ to penalize poor actions more strongly. In practice, gradients are detached from $w_i$ while flowing through $\hat{A}_i$. 
The resulting KL-weighted advantages $\hat{A}^{\mathrm{cap}}_{i}$ are substituted into the surrogate objective and combined with the caption consistency loss $\mathcal{L}_{\mathrm{cap}}$, producing group-level updates that favor perceptually consistent and correct trajectories.

\begin{algorithm}[t]
\caption{CapPO (modifications over GRPO in \textcolor{blue}{blue})}
\label{alg:method}
\begin{algorithmic}[1]
\vspace{+0.3em}
\STATE \textbf{Parameters}: MLLM policy $\pi_\theta$, dataset $\mathcal{D}$, group size $G$, clipping threshold $\epsilon$, \textcolor{blue}{caption-based consistency coefficient $\alpha$} and \textcolor{blue}{KL-weighted advantage coefficient $\beta$}.
\vspace{+0.5em}
\STATE \textbf{Dataset Construction:} Initialize dataset $\mathcal{D} = \{(i,q)\}$ with paired images and questions.
\FOR{each $(i,q)$ in $\mathcal{D}$}
    \STATE \textcolor{blue}{Generate caption $c = C(i,q)$ using external captioner}
    \STATE \textcolor{blue}{Augment dataset as $\mathcal{D} \leftarrow \mathcal{D} \cup \{(i,q,c)\}$}
\ENDFOR
\vspace{+0.5em}
\STATE \textbf{Training Loop:}
\FOR{each iteration}
    \STATE Sample triplet $(i,q,\color{blue}{c})$ from $\mathcal{D}$
    \STATE Generate group $\{o_1,\ldots,o_G\} \sim \pi_{\theta_{\text{old}}}(\cdot \mid i,q)$
    \STATE Evaluate rule-based outcome reward $r_i$ for each $o_i$
    \FOR{each update iteration}
        \STATE Compute group-relative advantage $\hat{A}_i$ by (\ref{eq:a})
        \STATE \textcolor{blue}{Compute caption-conditioned distribution $\pi_\theta(o \mid c,q)$}
        \vspace{-1.2em}
        \STATE \textcolor{blue}{Evaluate caption-based KL divergence $D_{\mathrm{KL}}(\pi_\theta^i \parallel \pi_\theta^c)$}
        
        \FOR{each response $o_i$}
            \STATE Compute importance sampling ratio $\rho_{i,t}$
            \STATE \textcolor{blue}{Aggregate $\bar{D}^{\mathrm{KL}}_i$ by (\ref{eq:seq-kl}) and normalize within group}
            \vspace{-1.2em}
            \STATE \textcolor{blue}{Compute  $w_i$ by (\ref{eq:weight}) and obtain $\hat{A}_i^{\mathrm{cap}}$ by (\ref{eq:weighted-adv})}
        \ENDFOR
        
        \STATE Compute overall objective $J_{\mathrm{CapPO}}(\theta)$ in (\ref{eq:overall})
        \STATE Update parameters $\theta \gets \theta + \eta \nabla_\theta J_{\mathrm{CapPO}}(\theta)$
    \ENDFOR
\ENDFOR
\end{algorithmic}
\end{algorithm}

\subsection{Overall Objective}
\label{sec:overall}


By integrating the standard GRPO objective with the proposed caption-based consistency regularization and KL-weighted advantage estimation, we obtain the overall training objective of CapPO, which can be formally expressed as
\begin{equation}
\label{eq:overall}
\begin{aligned}
&J_{\mathrm{CapPO}}(\theta)= J_{\mathrm{GRPO}}(\theta,{\textcolor{blue}{\hat{A}_i^{\mathrm{cap}}}}) - {\textcolor{blue}{\alpha \mathcal{L}_{\mathrm{cap}}}} \\
&= \mathbb{E}_{q,i,c, \{o_i\}_{i=1}^G} \Bigg[
\frac{1}{G} \sum_{i=1}^G \frac{1}{|o_i|} \sum_{t=1}^{|o_i|} 
\Big( \min\!\big( \rho_{i,t}{\textcolor{blue}{\hat{A}^{\mathrm{cap}}_{i}}}, \; \\
&\text{clip}(\rho_{i,t}, 1-\epsilon, 1+\epsilon){\textcolor{blue}{\hat{A}^{\mathrm{cap}}_{i}}}\big) 
  - {\textcolor{blue}{\alpha D_{\mathrm{KL}}\left(\pi_\theta^i  \parallel \pi_\theta^o\right)}} \Big) \Bigg],
\end{aligned}
\end{equation}
where $\alpha$ is a hyperparameter controlling the strength of the caption-based consistency regularization, and $D_{\mathrm{KL}}\!\left(\pi_\theta^i \parallel \pi_\theta^c\right)$ is a shorthand for $D_{\mathrm{KL}}\!\left(\pi_\theta(o_i \mid i, q)\, \parallel\, \pi_\theta(o_i \mid c, q)\right)$.
Following prior studies \cite{meng2025mm} and \cite{guo2025observe}, we remove the KL divergence loss with respect to $\pi_{\text{ref}}$.

The complete procedure is summarized in Algorithm~\ref{alg:method}, where modifications over GRPO are highlighted in {\textcolor{blue}{blue}}.

\section{Experiments}
\label{sec:experiments}

In this section, we aim to answer several key questions regarding the effectiveness of our proposed framework:
\begin{itemize}
    \item \textbf{Overall performance:} Does CapPO achieve competitive results across reasoning benchmarks?
    \item \textbf{Parameter sensitivity:} How sensitive is CapPO to the hyperparameters $\alpha$ and $\beta$?
    \item \textbf{Data efficiency and resources:} Can CapPO remain effective with reduced training data, and what is the additional computational overhead introduced by caption-based consistency regularization?
    \item \textbf{Error reduction:} Does CapPO reduce perception-related mistakes compared with existing baselines?
\end{itemize}

\subsection{Setup}
\subsubsection{Dataset}
For CapPO training, we adopt the \textit{ViRL39K} dataset introduced in prior work \cite{wang2025vl}. 
This dataset consists of approximately 38,870 multimodal reasoning questions spanning a diverse set of domains. 
The majority of the questions focus on mathematics, including geometric, non-geometric, and chart-diagrams problems (around 80\%), while the remaining questions come from general reasoning problems, including scientific, social and spatial understanding tasks (about 20\%). 
In addition, to obtain captions, we leverage the \textit{Qwen2.5-VL-72B} \cite{bai2025qwen2} to generate high-quality descriptions for all visual inputs. 
The prompt design strictly follows a structured template (see Fig.~\ref{fig_prompt} in the Appendix \ref{appendix:a} ), ensuring that the generated captions preserve both the semantic integrity and problem-solving relevance of the original questions. 
After filtering out samples with format inconsistencies or captioning failures, we retain a cleaned subset of about 36,579 valid questions, which forms the final dataset used in our experiments. This cleaned collection provides a reliable foundation for RL training with aligned visual and textual inputs.

\subsubsection{Baselines}
We build CapPO upon \textit{Qwen2.5-VL-7B} \cite{bai2025qwen2}, with the primary focus on multimodal reasoning, particularly tasks involving image-based reasoning. 
Accordingly, we compare against  open-source variants that are also based on \textit{Qwen2.5-VL-7B} but specifically trained on image-reasoning datasets. 
The considered baselines include OpenVL-Thinker~\cite{deng2025openvlthinker}, MM-Eureka~\cite{meng2025mm}, VL-Rethinker~\cite{wang2025vl}, Mixed-R1~\cite{xu2025mixed}, SophiaVL-R1~\cite{fan2025sophiavl}, R1-ShareVL~\cite{yao2025r1}, GThinker~\cite{zhan2025gthinker}, PAPO-G-NoRef~\cite{wang2025perception} and Shuffle-R1 \cite{zhu2025shuffle}. 

For fairness, we also compare the scale of SFT and RL data across these baselines, and note that\textbf{ CapPO achieves competitive performance while using a comparable or smaller amount of RL data (Table \ref{tab:overall_performance} and \ref{tab:baseline_data_scales})}.
Since our method is purely RL without any SFT, we mainly compare against baselines that are also trained with RL or with only light SFT on limited non–R1-style reasoning traces (i.e., without long iterative CoT). 
\textbf{All baselines were proposed in 2025 or later, with \textit{Shuffle-R1} (August 2025) \cite{zhu2025shuffle} being the most recent concurrent work}.

\begin{table*}[t]
\caption{Overall performance comparison of CapPO. Best and second results are highlighted in \textbf{bold} and \underline{underlined}, respectively.}
\label{tab:overall_performance}
\centering
\resizebox{\textwidth}{!}{%
\setlength{\tabcolsep}{2pt} 
\begin{tabular}{l ccccccc cccccc}
\toprule
\multirow{2}{*}{\textbf{Model}} 
& \multicolumn{6}{c}{\textbf{Math-Related Reasoning}} 
& \multicolumn{6}{c}{\textbf{General Reasoning}} \\
\cmidrule(lr){2-7} \cmidrule(lr){8-13}
& \textbf{MathVista} & \textbf{MathVerse} & \textbf{MathVision} & \textbf{WeMath} & \textbf{DynaMath} & \textbf{Avg} 
  & \textbf{LogicVista} & \textbf{ScienceQA} & \textbf{MMStar} & \textbf{MMMU} & \textbf{MMMU-Pro} & \textbf{Avg} \\
\midrule
Qwen2.5-VL-7B  & 69.2 & 44.7 & 21.7 & 34.8 & 53.7 & 44.8 & 46.9 & 88.8 & 64.6 & 56.3 & 40.8 & 59.5 \\
OpenVL-Thinker & 69.6 & 47.0 & 27.3 & 38.1 & 39.2 & 44.2 & 47.2 & 86.6 & 61.9 & \underline{59.4} & 41.3 & 59.3 \\
MM-Eureka      & 71.5 & \textbf{51.0} & 28.3 & 35.0 & 55.0 & 48.2 & 47.8 & 88.7 & 65.6 & 57.3 & 41.0 & 60.1 \\
Mixed-R1       & 70.2 & 46.7 & 27.0 & 37.2 & 54.1 & 47.0 & 46.8 & \underline{90.7} & 65.3 & 58.4 & 41.7 & 60.6 \\
SophiaVL-R1    & 70.4 & 44.8 & 27.6 & 39.6 & 54.5 & 47.4 & 46.3 & 90.1 & \underline{65.8} & 56.0 & 42.3 & 60.1 \\
VL-Rethinker   & 71.5 & \underline{50.9} & 28.5 & 42.0 & 56.0 & 49.8 & 43.6 & 87.6 & 64.7 & 59.2 & \underline{43.4} & 59.7 \\
R1-ShareVL     & \underline{74.7} & 44.9 & 29.3 & \textbf{43.0} & \underline{56.7} & 49.7 & \underline{49.1} & 89.9 & 64.5 & 59.3 & 42.8 & 61.1 \\
GThinker    & 69.8 & 47.9 & 27.6 & 40.4 & 54.6 & 48.1 & 45.6 & 90.6 & 64.6 & 58.1 & 42.0 & 60.2 \\
PAPO-G-NoRef   & 72.3 & 43.1 & 27.3 & 37.8 & 55.4 & 47.2 & 48.1 & 89.4 & 64.5 & 57.1 & 41.9 & 60.2 \\
Shuffle-R1  & \textbf{74.8} & 50.1 & \underline{30.0} & 40.1 & 55.2 & \underline{50.0} & \underline{49.1} & 90.5 & \textbf{66.7} & 58.5 & \underline{43.4} & \underline{61.6} \\
\textbf{CapPO (Ours)}   & 73.6 & 49.4 & \textbf{30.3} & \underline{42.6} & \textbf{57.9} & \textbf{50.8} & \textbf{49.4} & \textbf{91.0} & 64.8 & \textbf{60.0} & \textbf{44.1} & \textbf{61.9} \\
\rowcolor{cyan!10}
\textbf{$\Delta$(Qwen2.5-VL)}    & +4.4 & +4.7 & +8.6 & +7.8 & +4.2 & +6.0 & +2.5 & +2.2 & +0.2 & +3.7 & +3.3 & +2.4 \\
\bottomrule
\end{tabular}%
}
\vspace{-1em}
\end{table*}

\begin{table}[t]
\caption{Data scales of RL-based MLLM baselines.}
\label{tab:baseline_data_scales}
\centering
\resizebox{\columnwidth}{!}{
\setlength{\tabcolsep}{2.5pt}
\begin{tabular}{lcccl}
\toprule
\textbf{Model} & \textbf{Date(Y/M)} & \textbf{SFT} & \textbf{RL} & \textbf{Notes} \\
\midrule
Qwen2.5-VL-7B\cite{bai2025qwen2} & 25.2 & -  & -   & Base Model \\
OpenVL-Thinker\cite{deng2025openvlthinker} & 25.3 & \textcolor{green}{\ding{51}}  35K  & \textcolor{green}{\ding{51}} 15K   & Iterative SFT+RL \\
MM-Eureka\cite{meng2025mm}      & 25.3 & \textcolor{red}{\ding{55}}    & \textcolor{green}{\ding{51}} 54K   & MM-Eureka-54K \\
VL-Rethinker\cite{wang2025vl}   & 25.4 & \textcolor{red}{\ding{55}}    & \textcolor{green}{\ding{51}} 39K   & ViRL-39K \\
Mixed-R1\cite{yao2025r1}       & 25.5 & \textcolor{red}{\ding{55}}    & \textcolor{green}{\ding{51}} 45K   & Mixed-45K \\
SophiaVL-R1\cite{fan2025sophiavl}    & 25.5 & \textcolor{red}{\ding{55}}    & \textcolor{green}{\ding{51}} 130K  & SophiaVL-R1-130K \\
R1-ShareVL\cite{yao2025r1}     & 25.5 & \textcolor{red}{\ding{55}}    & \textcolor{green}{\ding{51}} 52K   & From MM-Eureka-54K \\
GThinker\cite{zhan2025gthinker} & 25.6 & \textcolor{green}{\ding{51}} 7K & \textcolor{green}{\ding{51}} 4K & GThinker-11K \\
PAPO-G-NoRef\cite{wang2025perception}   & 25.7 & \textcolor{red}{\ding{55}}    & \textcolor{green}{\ding{51}} 39K   & On ViRL-39K \\
Shuffle-R1\cite{zhu2025shuffle}   & 25.8 & \textcolor{red}{\ding{55}}    & \textcolor{green}{\ding{51}} 30K   & From MM-Eureka-54K \\
\textbf{CapPO (Ours)}   & -     & \textcolor{red}{\ding{55}}    & \textcolor{green}{\ding{51}} 36K   & From ViRL-39K \\
\bottomrule
\end{tabular}
}
\vspace{-1em}
\end{table}

\subsubsection{Benchmarks}
Since most of our training data comes from mathematical problems, we mainly emphasize evaluation on math-related benchmarks. 
To further assess generalization, we also include several general reasoning benchmarks. 
Overall, the evaluation is conducted on two categories: \textit{Math-Related Reasoning} (MathVista \cite{lu2023mathvista}, MathVerse \cite{zhang2024mathverse}, MathVision \cite{wang2024measuring}, WeMath \cite{qiao2024we}, and DynaMath \cite{zou2024dynamath}) and 
\textit{General Reasoning} (LogicVista \cite{xiao2024logicvista}, ScienceQA \cite{lu2022learn}, MMStar \cite{chen2024we}, MMMU \cite{yue2024mmmu}, and MMMU-Pro\cite{yue2024mmmupro}). 
The datasets mainly consist of several hundred to several thousand visual question answering (VQA) and multiple-choice question (MCQ) samples. 
More benchmark, training and evaluation details are provided in the Appendix \ref{appendix:b}.

\subsection{Overall Performance}
Table~\ref{tab:overall_performance} presents the overall comparison across math-related and general reasoning benchmarks. 
CapPO achieves competitive performance against the base model and all baselines, demonstrating consistent improvements across diverse evaluation settings.
The advantage is most pronounced on math-related benchmarks, which demand precise visual perception and multi-step reasoning, where CapPO shows clear superiority. 
Meanwhile, it also delivers competitive results on general reasoning tasks, confirming that the improvements generalize beyond a single domain. 
Overall, CapPO achieves state-of-the-art average accuracy, with gains of +6.0 on math-related and +2.4 on general reasoning tasks over the base model.

\begin{table}[t]
\caption{Effect of CapPO Components on Performance.}
\label{tab:ablation}
\centering
\setlength{\tabcolsep}{4pt}
\resizebox{1\columnwidth}{!}{%
\begin{tabular}{lcccccc}
\toprule
\textbf{Model} & \textbf{Math} & $\boldsymbol{\Delta}$ & \textbf{General} & $\boldsymbol{\Delta}$ & \textbf{Avg} & $\boldsymbol{\Delta}$ \\
\midrule
\rowcolor{yellow!20}
\textbf{Qwen2.5-VL-3B} & \textit{36.8} & - & \textit{52.1} & - & \textit{44.5} & - \\
+GRPO & 39.9 & \cellcolor{cyan!10}+3.1 & 54.1 & \cellcolor{cyan!10}+2.0 & 47.0 & \cellcolor{cyan!10}+2.5 \\
+CapPO w/o KL-Reg & 40.9 & \cellcolor{cyan!10}+4.1 & 54.6 & \cellcolor{cyan!10}+2.5 & 47.8 & \cellcolor{cyan!10}+3.3 \\
+CapPO w/o Adv-Weight & 40.7 & \cellcolor{cyan!10}+3.9 & 53.8 & \cellcolor{cyan!10}+1.7 & 47.3 & \cellcolor{cyan!10}+2.8 \\
\textbf{+CapPO} & \textbf{41.9} & \cellcolor{cyan!10}\textbf{+5.1} & \textbf{55.4} & \cellcolor{cyan!10}\textbf{+3.3} & \textbf{48.7} & \cellcolor{cyan!10}\textbf{+4.2} \\
\midrule
\rowcolor{yellow!20}
\textbf{Qwen2.5-VL-7B} & \textit{44.8} & - & \textit{59.5} & - & \textit{52.2} & - \\
+GRPO & 48.3 & \cellcolor{cyan!10}+3.5 & 61.0 & \cellcolor{cyan!10}+1.5 & 54.7 & \cellcolor{cyan!10}+2.3 \\
+CapPO w/o KL-Reg & 49.6 & \cellcolor{cyan!10}+4.8 & 60.3 & \cellcolor{cyan!10}+0.8 & 55.0 & \cellcolor{cyan!10}+2.6 \\
+CapPO w/o Adv-Weight & 49.5 & \cellcolor{cyan!10}+4.7 & 61.1 & \cellcolor{cyan!10}+1.6 & 55.3 & \cellcolor{cyan!10}+2.9 \\
\textbf{+CapPO} & \textbf{50.8} & \cellcolor{cyan!10}\textbf{+6.0} & \textbf{61.9} & \cellcolor{cyan!10}\textbf{+2.4} & \textbf{56.4} & \cellcolor{cyan!10}\textbf{+4.0} \\
\bottomrule
\end{tabular}%
}
\vspace{-1em}
\end{table}

\subsection{Ablation Study}

\subsubsection{Benchmark Performance}
To rigorously assess the contribution of each component in our framework, we conduct ablation experiments on both the 3B and 7B variants of Qwen2.5-VL, with results summarized in Table~\ref{tab:ablation}. Several insights emerge. First, introducing GRPO on top of the base model consistently improves performance across both math-related and general reasoning benchmarks, validating the effectiveness of RL with verifiable rewards. Beyond this, removing either the caption-based KL regularization (\textit{CapPO w/o KL-Reg}) or the KL-weighted advantage mechanism (\textit{CapPO w/o Adv-Weight}) still leads to notable gains over GRPO, confirming that each module independently enhances reasoning capability. Most importantly, integrating both modules yields the strongest results on both model scales, with average improvements of +4.2 on the 3B model and +4.0 on the 7B model, thereby demonstrating the complementary benefits of perceptual consistency regularization and adaptive advantage weighting.

\subsubsection{Training Dynamics}

\begin{figure}[t]
\centering
\includegraphics[width=\linewidth]{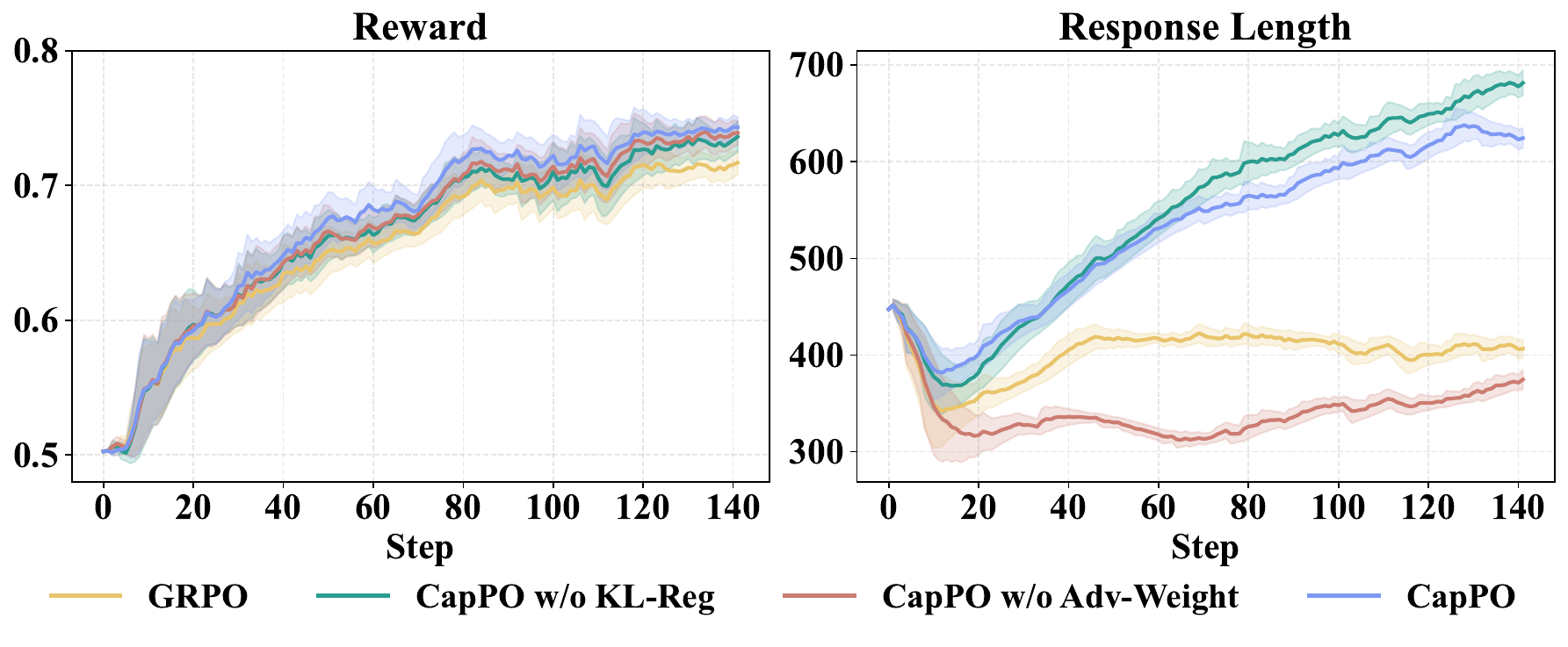}
\caption{Training rewards (left) and response lengths (right) of GRPO, CapPO and its variants on Qwen2.5-VL-7B.}
\label{fig:ablation}
\vspace{-1em}
\end{figure}

We further analyze the training dynamics on Qwen2.5-VL-7B in Fig.~\ref{fig:ablation}, which illustrates the trends of training reward and average response length. CapPO consistently achieves the highest and most stable reward throughout training. With respect to response length, CapPO shows an initial increase followed by stabilization.
In contrast, \emph{CapPO w/o KL-Reg} exhibits a monotonic increase in response length. We conjecture that this occurs because, with adaptive weighting, longer responses contain more reasoning content that aligns closely with the caption-conditioned logits, thereby receiving stronger advantages. Conversely, \emph{CapPO w/o Adv-Weight} shows a gradual reduction in response length, as longer responses generally accumulate larger KL across tokens and are consequently penalized more heavily.
These results highlight that both caption-based KL regularization and KL-weighted advantage estimation are indispensable for achieving stable and efficient reasoning.

\subsubsection{Hyperparameter Sensitivity}
\begin{figure}[t]
\centering
\includegraphics[width=\linewidth]{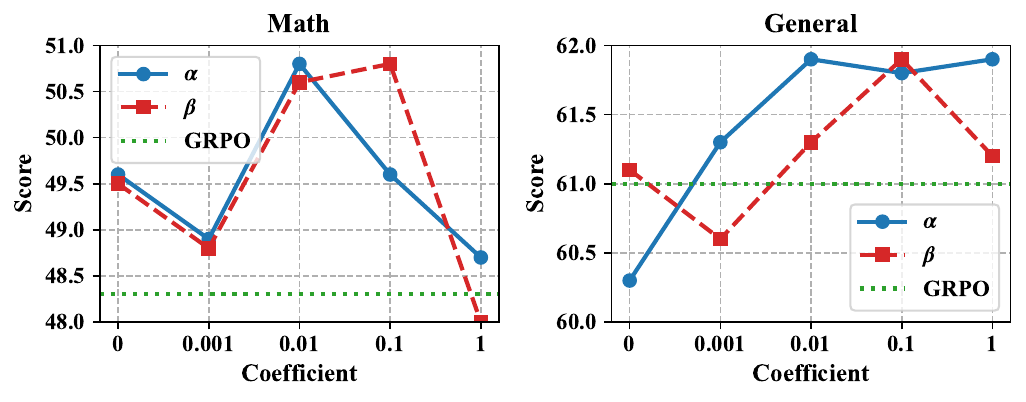}
\caption{CapPO performance sensitivity to hyperparameters $\alpha$ and $\beta$ on Math and General benchmarks based on Qwen2.5-VL-7B.}
\label{fig:ablation_ab}
\vspace{-1em}
\end{figure}

We further analyze the sensitivity of CapPO to the hyperparameters $\alpha$ (caption-based KL regularization) and $\beta$ (KL-weighted advantage). 
When varying $\alpha$, we fix $\beta=0.1$; conversely, when varying $\beta$,  we fix $\alpha=0.01$. The results in Fig.~\ref{fig:ablation_ab} show that  $\alpha=0.01$ and $\beta=0.1$ achieve the best performance on both Math and General benchmarks. 
Notably, across most small values of $\alpha$ and $\beta$,  CapPO consistently outperforms GRPO, which confirms the effectiveness of the  two proposed modules. On the other hand, excessively large coefficients, such as $1.0$, lead to clear performance degradation, indicating that overly strong  regularization can harm optimization by over-constraining the policy.
Due to computational resource limits, we did not conduct a more exhaustive search,  but these results suggest that parameters in this order of magnitude strike a favorable balance between stability and effectiveness.

\begin{figure}[t]
\centering
\includegraphics[width=\linewidth]{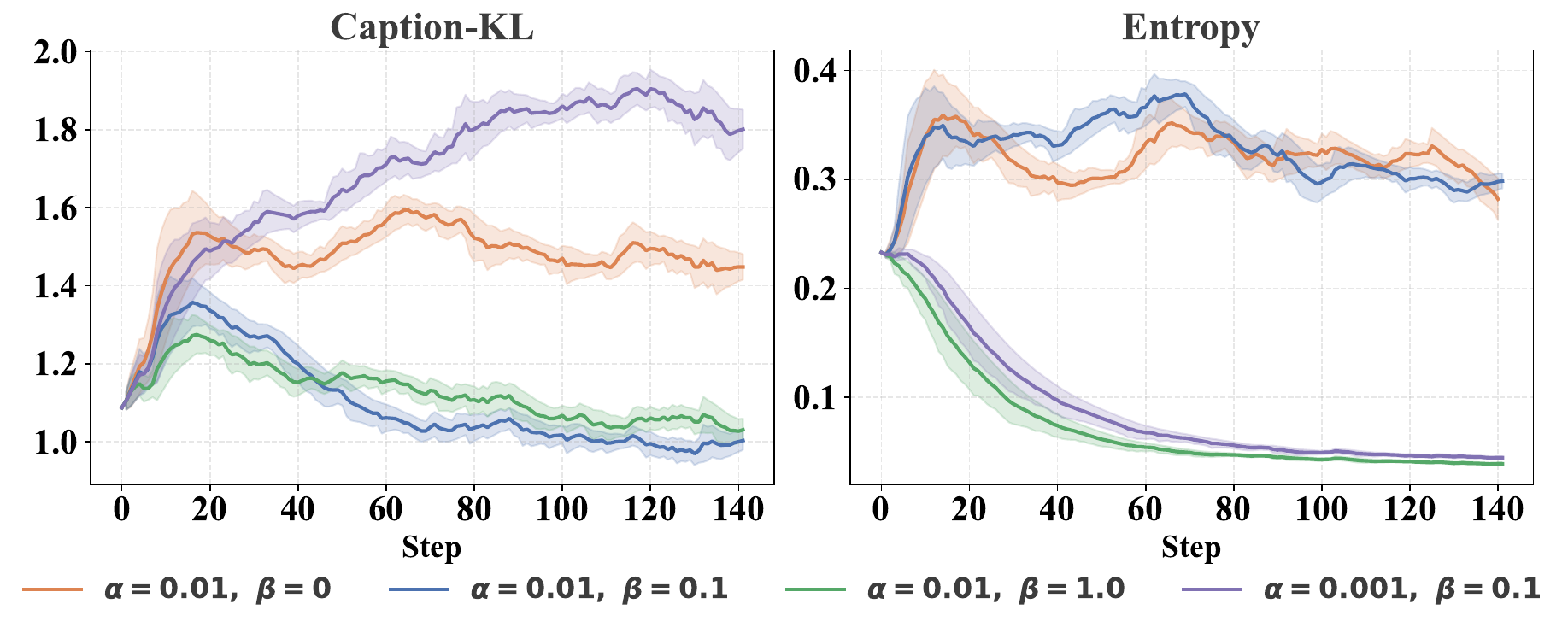}
\caption{Effects of $\alpha$ and $\beta$ on training caption KL (left) and entropy (right). Moderate values ($\alpha=0.01$, $\beta=0.1$) yield stable KL reduction and balanced entropy, while overly small or large coefficients lead to inferior behavior.}
\label{fig:ablation_klentropy}
\vspace{-1em}
\end{figure}

To further understand the role of the two coefficients during training, we analyze their effects on both caption-KL loss and entropy, as shown in Fig.~\ref{fig:ablation_klentropy}. 
A smaller caption-KL indicates that the output distribution of the model is becoming closer between caption-augmented questions and image-based questions, which aligns with our expectation that the model should produce consistent predictions under 
different perceptual conditions. 
Entropy, on the other hand, measures the degree of exploration in the policy and is defined as 
$H(\pi) = - \sum_{a} \pi(a | s) \log \pi(a | s)$. 
If entropy drops too quickly, the model converges prematurely and loses exploration ability, whereas moderate entropy helps maintain diversity during training and avoids suboptimal local minima \cite{cui2025entropy,fu2025srft}.

We observe that both $\alpha$ and $\beta$ play important roles in constraining the caption-KL loss. When these coefficients are too small (the purple and blue curves), the KL constraint cannot be effectively enforced, leading to suboptimal perceptual alignment. 
For entropy, an 
overly large $\beta$ drives the policy toward excessive determinism (green curve), likely because the advantage of certain reasoning patterns is over-amplified. Conversely, extremely small entropy (purple curve) can also hinder exploration, which can be explained by the fact that the caption-KL constraint implicitly encourages diversity in addition to optimizing correctness in the original RL objective. 
Taken together, these observations indicate that $\alpha$ and $\beta$ strongly influence the reasoning patterns learned by the model. 
We therefore recommend setting $\alpha=0.01$ and $\beta=0.1$ (orange curve), which yield stable KL reduction and moderate entropy throughout training.

\subsubsection{Captioner Ablation}
In the previous experiments, captions were obtained from a stronger external model, which provided reliable semantic anchors for enforcing perceptual consistency. 
A natural question is whether CapPO can still outperform GRPO when such a strong captioner is not available. 
To investigate this, we conduct an ablation study where the training captions are generated by the same \textit{Qwen2.5-VL} model used for policy learning (7B) or even a smaller variant (3B). 
Table \ref{tab:ablation_caption} reports the effect of using different captioners in CapPO. 
Compared with GRPO, all variants of CapPO consistently improve average accuracy. 
In particular, a stronger captioner (72B) brings the largest gain (+1.7), while even the self-generated captions from the 7B model (+1.3) or a smaller 3B variant (+0.9) still outperform GRPO. 
These findings demonstrate two important points: (1) stronger captioners provide more reliable semantic anchors, leading to larger performance gains; (2) Even when only weaker or self-generated captions are available, CapPO still surpasses GRPO, indicating that the framework is less sensitive to caption quality and remains beneficial in low-resource scenarios.

\begin{table}[!t]
\caption{Performance with different captioners for CapPO. }
\label{tab:ablation_caption}
\centering
\setlength{\tabcolsep}{4pt}
\resizebox{.9\columnwidth}{!}{%
\begin{tabular}{lcccccc}
\toprule
\textbf{Captioner} & \textbf{Math} & $\boldsymbol{\Delta}$ & \textbf{General} & $\boldsymbol{\Delta}$ & \textbf{Avg} & $\boldsymbol{\Delta}$ \\
\midrule
\rowcolor{yellow!20}
\textbf{- (GRPO)} & \textit{48.3} & - & \textit{61.0} & - & \textit{54.7} & - \\
\textbf{3B (Smaller)} & 50.5 & \cellcolor{cyan!10}+2.2 & 60.7 & \cellcolor{cyan!10}-0.3 & 55.6 & \cellcolor{cyan!10}+0.9 \\
\textbf{7B (Self)} & 50.7 & \cellcolor{cyan!10}+2.4 & 61.2 & \cellcolor{cyan!10}+0.2 & 56.0 & \cellcolor{cyan!10}+1.3 \\
\textbf{72B (Larger)} & 50.8 & \cellcolor{cyan!10}+2.5 & 61.9 & \cellcolor{cyan!10}+0.9 & 56.4 & \cellcolor{cyan!10}+1.7 \\
\bottomrule
\end{tabular}%
}
\vspace{-1em}
\end{table}

\subsubsection{Dataset Scaling}
\begin{figure}[t]
\centering
\includegraphics[width=\linewidth]{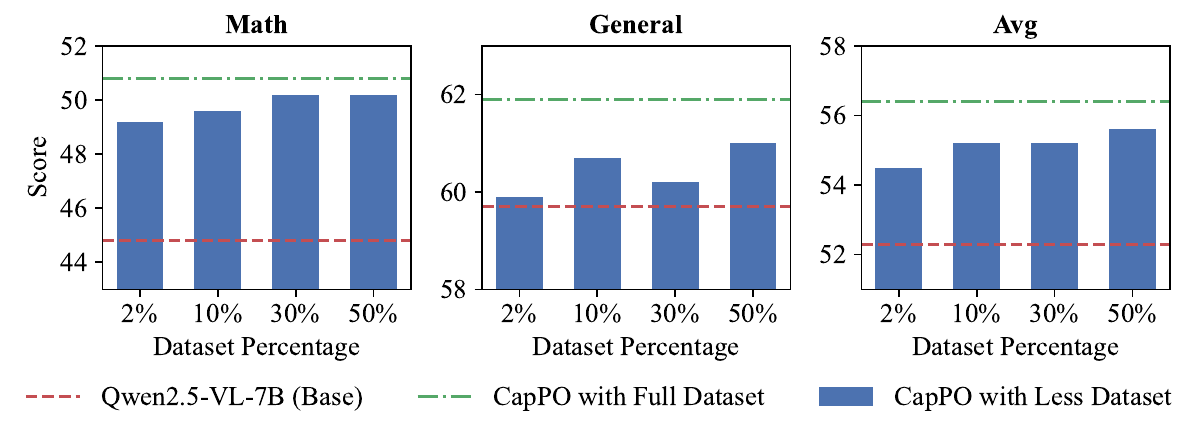}
\caption{
Ablation on dataset scaling. Performance drops as data decreases, but even with only 2\% of the data, CapPO outperforms the base model.
}
\vspace{-1em}
\label{fig_dataset_ablation}
\end{figure}

To examine the effect of training data scale, we randomly subsample different proportions of the training dataset. For the 30\% and 50\% settings, we train for the same number of 140 steps as the full-data setup to ensure fair comparison. For the 2\% and 10\% subsets, we adjust training length to 40 and 70 steps, respectively, based on validation performance. As shown in Fig.~\ref{fig_dataset_ablation}, performance decreases gradually as the dataset size is reduced. 
Nevertheless, even with only 2\% of the data (732 samples), CapPO training achieves higher accuracy than the base model, demonstrating the effectiveness of our approach under limited data regimes.

\subsection{Analysis}

\subsubsection{Error Distributions}
\begin{figure}[t]
\centering
\includegraphics[width=\linewidth]{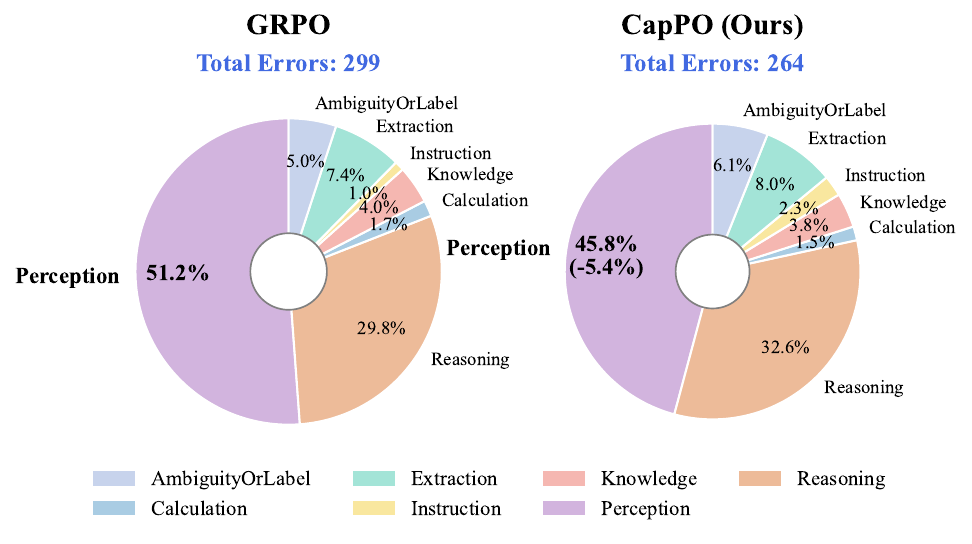}
\caption{Error-type distributions across different models on MathVista. Each pie chart shows the relative proportions of major error categories: \textit{Perception} (misread input, tables, or figures), \textit{Reasoning} (flawed or incomplete logical chain), \textit{Calculation} (arithmetic or unit conversion mistakes), \textit{Knowledge} (incorrect facts or definitions), \textit{Instruction} (failure to follow explicit instructions or output format), \textit{Extraction} (wrong final answer extraction despite correct reasoning), and \textit{AmbiguityOrLabel} (ambiguous questions or suspicious gold labels). CapPO notably reduces perception errors (-5.4\%) compared to GRPO.}
\label{fig_pies}
\vspace{-1em}
\end{figure}

To diagnose failure modes, we analyze mispredictions on the \textit{MathVista test\_mini} set for three models: Qwen2.5-VL-7B (base), GRPO, and CapPO. 
For each \textit{(question, response)} pair, we prompt GPT-5 to assign one of seven error categories (\textit{Perception, Reasoning, Calculation, Knowledge, Instruction, Extraction, AmbiguityOrLabel}), as detailed in Fig.~\ref{fig_pies}. 
The base model yields 308 total errors of total 1,000 samples, with \textit{Perception} accounting for 51.3\%, as show in Fig.~\ref{fig_qwen_error}. GRPO reduces the total to 299, but the proportion of perception errors remains essentially unchanged, suggesting that its gains do not primarily address perceptual failures. 
By contrast, CapPO further decreases the total to 264 and substantially lowers the perception-error share to 45.8\%, a 5.4 percentage point reduction compared to GRPO, indicating that CapPO’s improvements are driven by suppressing perception-related mistakes and strengthening perceptual grounding for subsequent reasoning.

\subsubsection{Case Study}
\begin{figure*}[t]
\centering
\includegraphics[width=\textwidth]{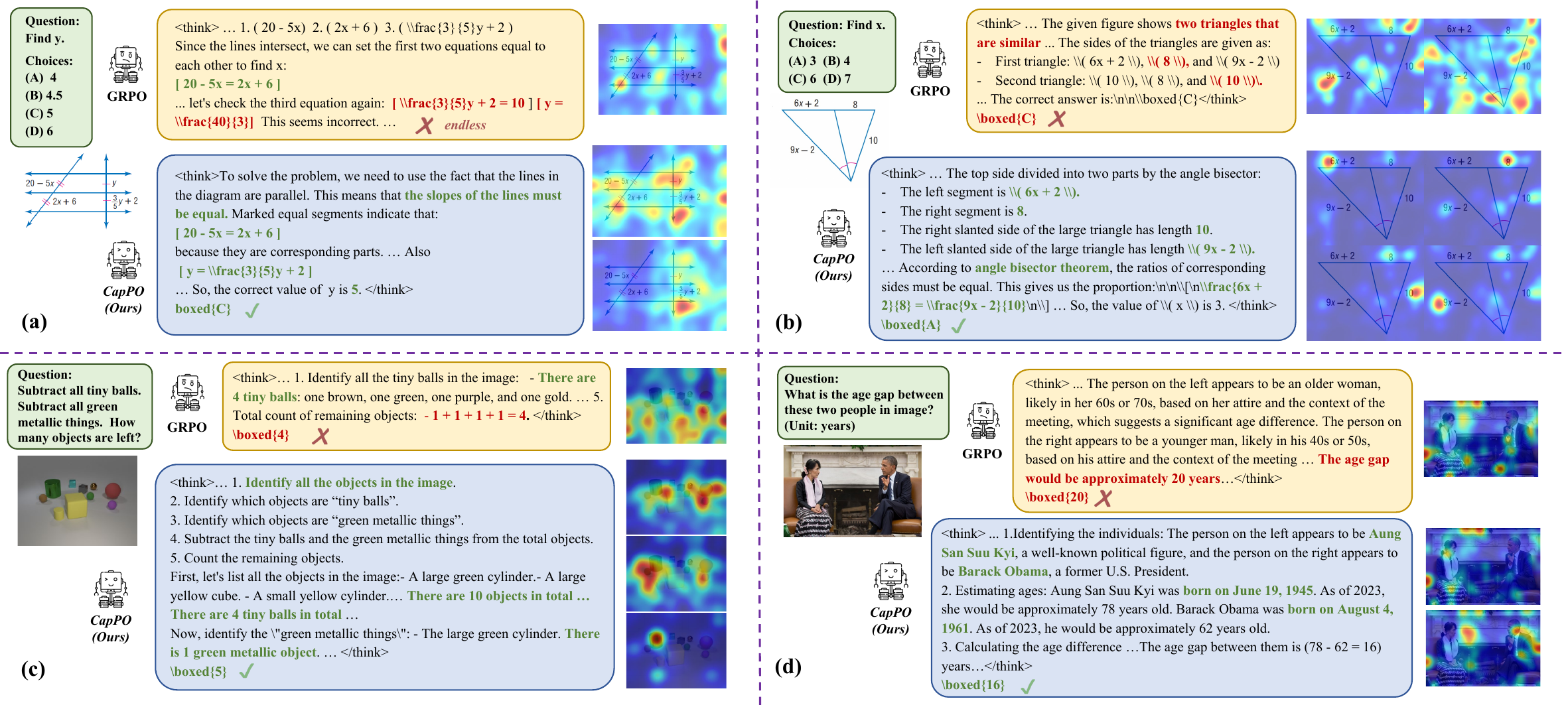}
\caption{
Case studies comparing GRPO and CapPO, with attention maps from the last 8 layers visualized. 
(a) Geometry reasoning: GRPO attempted to equate unrelated expressions, leading to inconsistent or endless reasoning, and its attention was scattered away from the critical parallel line segments. In contrast, CapPO explicitly exploited the parallelism, focused attention on the equal-length parts of the diagram, and correctly solved $y=5$. 
(b) Symbolic math: GRPO prematurely fixed inconsistent equations and failed to apply the parallelism constraint, with diffused attention across irrelevant symbols. CapPO localized its attention on the relevant numeric tokens and equal segments, derived consistent equations, and reached the correct solution. 
(c) Object counting: GRPO identified the four tiny balls but ignored the green metallic object, giving an undercount. CapPO systematically enumerated all objects, separated tiny balls and metallic objects, and kept strong attention on both the spheres and the green cylinder, arriving at the correct answer of 5. 
(d) Age estimation: GRPO relied on heuristic visual cues and guessed a 20-year gap, with attention concentrated on surrounding text rather than numbers. CapPO correctly identified Aung San Suu Kyi and Barack Obama, retrieved their birth years, and performed subtraction to obtain the correct 16-year difference, with attention sharply aligned to the numeric entities. 
}
\label{fig_case_studies}
\vspace{-1em}
\end{figure*}

To further illustrate the advantages of CapPO over baseline methods, we conduct a series of representative case studies comparing the reasoning trajectories and attention visualizations of GRPO and CapPO, as shown in Fig.~\ref{fig_case_studies}. 
For each example, we present the attention distributions from the last eight transformer layers, which clearly illustrate where the models allocate their focus during the reasoning process.
Across all four cases, GRPO exhibits several characteristic failure modes: 
(a) misapplication of algebraic rules leading to inconsistent derivations, 
(b) incomplete enumeration in object-centric visual tasks, 
(c) neglect of task-specific structural constraints, and 
(d) reliance on heuristic guesses in open-domain scenarios. 
In contrast, CapPO combines perceptual grounding with symbolic reasoning and maintains focused attention on task-relevant regions, thereby improving both accuracy and interpretability while avoiding the heuristic shortcuts and incomplete logic often seen in GRPO.

\subsubsection{Computational Cost}
CapPO introduces only minor additional overhead compared with GRPO, and the overall training cost remains of the same order of magnitude, requiring about 14 hours on an 8$\times$H100 GPU cluster versus roughly 11 hours for GRPO. 
See Appendix~\ref{appendix:train_cost} owing to space constraints.

\section{Conclusion}
In this paper, we proposed Caption-Regularized Policy Optimization (CapPO), a RL framework designed to enhance the reliability of multimodal reasoning. 
By introducing caption-based consistency regularization and KL-weighted advantage estimation, CapPO effectively reduces perception-induced errors and strengthens the alignment between visual grounding and symbolic reasoning. 
Extensive experiments across five math-focused and five general benchmarks demonstrated that CapPO achieves consistent improvements over GRPO and other competitive baselines, with particularly notable gains on perception-intensive tasks. 
Ablation studies and error analyses further verified the complementary roles of the two proposed modules, highlighting that perceptual consistency is a critical factor for multimodal reasoning.

Despite these promising results, several aspects remain open for future exploration. 
First, the current method relies on manual hyperparameter tuning, and developing automatic strategies to adaptively determine the coefficients could make the framework more reliable and easier to deploy in practice.
Second, CapPO is built on a zero-shot RL paradigm; extending it to hybrid approaches that combine SFT with RL, especially in settings involving interleaved vision–language reasoning, would further broaden its applicability. 
Finally, although CapPO already demonstrates substantial reductions in perception errors, it is worth emphasizing that our framework is largely orthogonal to other recent advances such as DAPO \cite{yu2025dapo} and Shuffle-R1 \cite{zhu2025shuffle}. Consequently, future work may explore integrating CapPO with these approaches to further enhance reasoning performance.
These directions may further advance perception-enhanced multimodal RL.

\bibliographystyle{IEEEtran}
\bibliography{refs}

\begin{thebibliography}{10}
\providecommand{\url}[1]{#1}
\csname url@samestyle\endcsname
\providecommand{\newblock}{\relax}
\providecommand{\bibinfo}[2]{#2}
\providecommand{\BIBentrySTDinterwordspacing}{\spaceskip=0pt\relax}
\providecommand{\BIBentryALTinterwordstretchfactor}{4}
\providecommand{\BIBentryALTinterwordspacing}{\spaceskip=\fontdimen2\font plus
\BIBentryALTinterwordstretchfactor\fontdimen3\font minus \fontdimen4\font\relax}
\providecommand{\BIBforeignlanguage}[2]{{%
\expandafter\ifx\csname l@#1\endcsname\relax
\typeout{** WARNING: IEEEtran.bst: No hyphenation pattern has been}%
\typeout{** loaded for the language `#1'. Using the pattern for}%
\typeout{** the default language instead.}%
\else
\language=\csname l@#1\endcsname
\fi
#2}}
\providecommand{\BIBdecl}{\relax}
\BIBdecl

\bibitem{liu2025reasonplan}
X.~Liu, Z.~Zhong, Y.~Guo, Y.~Liu, Z.~Su, Q.~Zhang, J.~Wang, Y.~Gao, Y.~Zheng, Q.~Lin \emph{et~al.}, ``Reasonplan: Unified scene prediction and decision reasoning for closed-loop autonomous driving,'' in \emph{Proceedings of the Conference on Robot Learning (CoRL)}, 2025.

\bibitem{zheng2024planagent}
Y.~Zheng, Z.~Xing, Q.~Zhang, B.~Jin, P.~Li, Y.~Zheng, Z.~Xia, K.~Zhan, X.~Lang, Y.~Chen \emph{et~al.}, ``Planagent: A multi-modal large language agent for closed-loop vehicle motion planning,'' \emph{arXiv preprint arXiv:2406.01587}, 2024.

\bibitem{li2024manipllm}
X.~Li, M.~Zhang, Y.~Geng, H.~Geng, Y.~Long, Y.~Shen, R.~Zhang, J.~Liu, and H.~Dong, ``Manipllm: Embodied multimodal large language model for object-centric robotic manipulation,'' in \emph{Proceedings of the IEEE/CVF Conference on Computer Vision and Pattern Recognition(CVPR)}, 2024, pp. 18\,061--18\,070.

\bibitem{chen2025robogpt}
Y.~Chen, W.~Cui, Y.~Chen, M.~Tan, X.~Zhang, J.~Liu, H.~Li, D.~Zhao, and H.~Wang, ``Robogpt: an llm-based long-term decision-making embodied agent for instruction following tasks,'' \emph{IEEE Transactions on Cognitive and Developmental Systems}, 2025.

\bibitem{liu2023multi}
Z.~Liu, J.~Cheng, J.~Fan, S.~Lin, Y.~Wang, and X.~Zhao, ``Multi-modal fusion based on depth adaptive mechanism for 3d object detection,'' \emph{IEEE Transactions on Multimedia}, vol.~27, pp. 707--717, 2023.

\bibitem{xiong20253ur}
H.~Xiong, Y.~Zhuge, J.~Zhu, L.~Zhang, and H.~Lu, ``3ur-llm: An end-to-end multimodal large language model for 3d scene understanding,'' \emph{IEEE Transactions on Multimedia}, 2025.

\bibitem{guo2025deepseek}
D.~Guo, D.~Yang, H.~Zhang, J.~Song, R.~Zhang, R.~Xu, Q.~Zhu, S.~Ma, P.~Wang, X.~Bi \emph{et~al.}, ``Deepseek-r1: Incentivizing reasoning capability in llms via reinforcement learning,'' \emph{arXiv preprint arXiv:2501.12948}, 2025.

\bibitem{tu2025learning}
S.~Tu, J.~Lin, Q.~Zhang, X.~Tian, L.~Li, X.~Lan, and D.~Zhao, ``Learning when to think: Shaping adaptive reasoning in r1-style models via multi-stage rl,'' \emph{Advances in Neural Information Processing Systems (NeurIPS)}, 2025.

\bibitem{fu2025rlae}
Y.~Fu, Y.~Zhu, J.~Chai, G.~Yin, W.~Lin, Q.~Zhang, and D.~Zhao, ``Rlae: Reinforcement learning-assisted ensemble for llms,'' in \emph{Proceedings of the Conference on Empirical Methods in Natural Language Processing (EMNLP)}, 2025.

\bibitem{xie2025logic}
T.~Xie, Z.~Gao, Q.~Ren, H.~Luo, Y.~Hong, B.~Dai, J.~Zhou, K.~Qiu, Z.~Wu, and C.~Luo, ``Logic-rl: Unleashing llm reasoning with rule-based reinforcement learning,'' \emph{arXiv preprint arXiv:2502.14768}, 2025.

\bibitem{deng2025openvlthinker}
Y.~Deng, H.~Bansal, F.~Yin, N.~Peng, W.~Wang, and K.-W. Chang, ``Openvlthinker: An early exploration to complex vision-language reasoning via iterative self-improvement,'' \emph{arXiv preprint arXiv:2503.17352}, 2025.

\bibitem{meng2025mm}
F.~Meng, L.~Du, Z.~Liu, Z.~Zhou, Q.~Lu, D.~Fu, T.~Han, B.~Shi, W.~Wang, J.~He \emph{et~al.}, ``Mm-eureka: Exploring the frontiers of multimodal reasoning with rule-based reinforcement learning,'' \emph{arXiv preprint arXiv:2503.07365}, 2025.

\bibitem{wang2025vl}
H.~Wang, C.~Qu, Z.~Huang, W.~Chu, F.~Lin, and W.~Chen, ``Vl-rethinker: Incentivizing self-reflection of vision-language models with reinforcement learning,'' \emph{arXiv preprint arXiv:2504.08837}, 2025.

\bibitem{liu2024survey}
H.~Liu, W.~Xue, Y.~Chen, D.~Chen, X.~Zhao, K.~Wang, L.~Hou, R.~Li, and W.~Peng, ``A survey on hallucination in large vision-language models,'' \emph{arXiv preprint arXiv:2402.00253}, 2024.

\bibitem{zhou2025reinforced}
G.~Zhou, P.~Qiu, C.~Chen, J.~Wang, Z.~Yang, J.~Xu, and M.~Qiu, ``Reinforced mllm: A survey on rl-based reasoning in multimodal large language models,'' \emph{arXiv preprint arXiv:2504.21277}, 2025.

\bibitem{wang2025perception}
Z.~Wang, X.~Guo, S.~Stoica, H.~Xu, H.~Wang, H.~Ha, X.~Chen, Y.~Chen, M.~Yan, F.~Huang \emph{et~al.}, ``Perception-aware policy optimization for multimodal reasoning,'' \emph{arXiv preprint arXiv:2507.06448}, 2025.

\bibitem{li2025vision}
Y.~Li, L.~Wei, K.~Zheng, J.~Huang, L.~Kong, L.~Sun, and W.~Huang, ``Vision matters: Simple visual perturbations can boost multimodal math reasoning,'' \emph{arXiv preprint arXiv:2506.09736}, 2025.

\bibitem{lu2023mathvista}
P.~Lu, H.~Bansal, T.~Xia, J.~Liu, C.~Li, H.~Hajishirzi, H.~Cheng, K.-W. Chang, M.~Galley, and J.~Gao, ``Mathvista: Evaluating mathematical reasoning of foundation models in visual contexts,'' in \emph{International Conference on Learning Representations (ICLR)}, 2024.

\bibitem{yue2024mmmu}
X.~Yue, Y.~Ni, K.~Zhang, T.~Zheng, R.~Liu, G.~Zhang, S.~Stevens, D.~Jiang, W.~Ren, Y.~Sun \emph{et~al.}, ``Mmmu: A massive multi-discipline multimodal understanding and reasoning benchmark for expert agi,'' in \emph{Proceedings of the IEEE/CVF Conference on Computer Vision and Pattern Recognition (CVPR)}, 2024, pp. 9556--9567.

\bibitem{xu2025mixed}
S.~Xu, Y.~Li, R.~Yang, T.~Zhang, Y.~Sun, W.~Chow, L.~Li, H.~Song, Q.~Xu, Y.~Tong \emph{et~al.}, ``Mixed-r1: Unified reward perspective for reasoning capability in multimodal large language models,'' \emph{arXiv preprint arXiv:2505.24164}, 2025.

\bibitem{fan2025sophiavl}
K.~Fan, K.~Feng, H.~Lyu, D.~Zhou, and X.~Yue, ``Sophiavl-r1: Reinforcing mllms reasoning with thinking reward,'' \emph{arXiv preprint arXiv:2505.17018}, 2025.

\bibitem{yao2025r1}
H.~Yao, Q.~Yin, J.~Zhang, M.~Yang, Y.~Wang, W.~Wu, F.~Su, L.~Shen, M.~Qiu, D.~Tao \emph{et~al.}, ``R1-sharevl: Incentivizing reasoning capability of multimodal large language models via share-grpo,'' \emph{arXiv preprint arXiv:2505.16673}, 2025.

\bibitem{zhan2025gthinker}
Y.~Zhan, Z.~Wu, Y.~Zhu, R.~Xue, R.~Luo, Z.~Chen, C.~Zhang, Y.~Li, Z.~He, Z.~Yang \emph{et~al.}, ``Gthinker: Towards general multimodal reasoning via cue-guided rethinking,'' \emph{arXiv preprint arXiv:2506.01078}, 2025.

\bibitem{huang2025vision}
W.~Huang, B.~Jia, Z.~Zhai, S.~Cao, Z.~Ye, F.~Zhao, Z.~Xu, Y.~Hu, and S.~Lin, ``Vision-r1: Incentivizing reasoning capability in multimodal large language models,'' \emph{arXiv preprint arXiv:2503.06749}, 2025.

\bibitem{zhang2025r1}
J.~Zhang, J.~Huang, H.~Yao, S.~Liu, X.~Zhang, S.~Lu, and D.~Tao, ``R1-vl: Learning to reason with multimodal large language models via step-wise group relative policy optimization,'' \emph{arXiv preprint arXiv:2503.12937}, 2025.

\bibitem{zhu2025shuffle}
L.~Zhu, Y.~Guan, D.~Liang, J.~Ju, Z.~Luo, B.~Qin, J.~Luan, Y.~Liu, and X.~Bai, ``Shuffle-r1: Efficient rl framework for multimodal large language models via data-centric dynamic shuffle,'' \emph{arXiv preprint arXiv:2508.05612}, 2025.

\bibitem{liu2022depth}
Y.~Liu, W.~Wei, D.~Peng, X.-L. Mao, Z.~He, and P.~Zhou, ``Depth-aware and semantic guided relational attention network for visual question answering,'' \emph{IEEE Transactions on Multimedia}, vol.~25, pp. 5344--5357, 2022.

\bibitem{ma2023boosting}
G.~Ma, Y.~Bai, W.~Zhang, T.~Yao, B.~Shihada, and T.~Mei, ``Boosting generic visual-linguistic representation with dynamic contexts,'' \emph{IEEE Transactions on Multimedia}, vol.~25, pp. 8445--8457, 2023.

\bibitem{xiao2023clip}
L.~Xiao, X.~Yang, F.~Peng, M.~Yan, Y.~Wang, and C.~Xu, ``Clip-vg: Self-paced curriculum adapting of clip for visual grounding,'' \emph{IEEE Transactions on Multimedia}, vol.~26, pp. 4334--4347, 2023.

\bibitem{radford2021learning}
A.~Radford, J.~W. Kim, C.~Hallacy, A.~Ramesh, G.~Goh, S.~Agarwal, G.~Sastry, A.~Askell, P.~Mishkin, J.~Clark \emph{et~al.}, ``Learning transferable visual models from natural language supervision,'' in \emph{International conference on machine learning (ICML)}.\hskip 1em plus 0.5em minus 0.4em\relax PmLR, 2021, pp. 8748--8763.

\bibitem{jia2021scaling}
C.~Jia, Y.~Yang, Y.~Xia, Y.-T. Chen, Z.~Parekh, H.~Pham, Q.~Le, Y.-H. Sung, Z.~Li, and T.~Duerig, ``Scaling up visual and vision-language representation learning with noisy text supervision,'' in \emph{International conference on machine learning (ICML)}.\hskip 1em plus 0.5em minus 0.4em\relax PMLR, 2021, pp. 4904--4916.

\bibitem{tschannen2025siglip}
M.~Tschannen, A.~Gritsenko, X.~Wang, M.~F. Naeem, I.~Alabdulmohsin, N.~Parthasarathy, T.~Evans, L.~Beyer, Y.~Xia, B.~Mustafa \emph{et~al.}, ``Siglip 2: Multilingual vision-language encoders with improved semantic understanding, localization, and dense features,'' \emph{arXiv preprint arXiv:2502.14786}, 2025.

\bibitem{fan2025grit}
Y.~Fan, X.~He, D.~Yang, K.~Zheng, C.-C. Kuo, Y.~Zheng, S.~J. Narayanaraju, X.~Guan, and X.~E. Wang, ``Grit: Teaching mllms to think with images,'' \emph{arXiv preprint arXiv:2505.15879}, 2025.

\bibitem{wu2025grounded}
Q.~Wu, X.~Yang, Y.~Zhou, C.~Fang, B.~Song, X.~Sun, and R.~Ji, ``Grounded chain-of-thought for multimodal large language models,'' \emph{arXiv preprint arXiv:2503.12799}, 2025.

\bibitem{xiao2025fast}
W.~Xiao, L.~Gan, W.~Dai, W.~He, Z.~Huang, H.~Li, F.~Shu, Z.~Yu, P.~Zhang, H.~Jiang \emph{et~al.}, ``Fast-slow thinking for large vision-language model reasoning,'' \emph{arXiv preprint arXiv:2504.18458}, 2025.

\bibitem{chen2025grpo}
Y.~Chen, Y.~Ge, R.~Wang, Y.~Ge, J.~Cheng, Y.~Shan, and X.~Liu, ``Grpo-care: Consistency-aware reinforcement learning for multimodal reasoning,'' \emph{arXiv preprint arXiv:2506.16141}, 2025.

\bibitem{yu2025perception}
E.~Yu, K.~Lin, L.~Zhao, J.~Yin, Y.~Wei, Y.~Peng, H.~Wei, J.~Sun, C.~Han, Z.~Ge \emph{et~al.}, ``Perception-r1: Pioneering perception policy with reinforcement learning,'' \emph{arXiv preprint arXiv:2504.07954}, 2025.

\bibitem{bai2025qwen2}
S.~Bai, K.~Chen, X.~Liu, J.~Wang, W.~Ge, S.~Song, K.~Dang, P.~Wang, S.~Wang, J.~Tang \emph{et~al.}, ``Qwen2. 5-vl technical report,'' \emph{arXiv preprint arXiv:2502.13923}, 2025.

\bibitem{guo2025observe}
Z.~Guo, M.~Hong, and T.~Jin, ``Observe-r1: Unlocking reasoning abilities of mllms with dynamic progressive reinforcement learning,'' \emph{arXiv preprint arXiv:2505.12432}, 2025.

\bibitem{zhang2024mathverse}
R.~Zhang, D.~Jiang, Y.~Zhang, H.~Lin, Z.~Guo, P.~Qiu, A.~Zhou, P.~Lu, K.-W. Chang, Y.~Qiao \emph{et~al.}, ``Mathverse: Does your multi-modal llm truly see the diagrams in visual math problems?'' in \emph{European Conference on Computer Vision (ECCV)}.\hskip 1em plus 0.5em minus 0.4em\relax Springer, 2024, pp. 169--186.

\bibitem{wang2024measuring}
K.~Wang, J.~Pan, W.~Shi, Z.~Lu, H.~Ren, A.~Zhou, M.~Zhan, and H.~Li, ``Measuring multimodal mathematical reasoning with math-vision dataset,'' \emph{Advances in Neural Information Processing Systems (NeurIPS)}, vol.~37, pp. 95\,095--95\,169, 2024.

\bibitem{qiao2024we}
R.~Qiao, Q.~Tan, G.~Dong, M.~Wu, C.~Sun, X.~Song, Z.~GongQue, S.~Lei, Z.~Wei, M.~Zhang \emph{et~al.}, ``We-math: Does your large multimodal model achieve human-like mathematical reasoning?'' \emph{arXiv preprint arXiv:2407.01284}, 2024.

\bibitem{zou2024dynamath}
C.~Zou, X.~Guo, R.~Yang, J.~Zhang, B.~Hu, and H.~Zhang, ``Dynamath: A dynamic visual benchmark for evaluating mathematical reasoning robustness of vision language models,'' in \emph{International Conference on Learning Representations (ICLR)}, 2025.

\bibitem{xiao2024logicvista}
Y.~Xiao, E.~Sun, T.~Liu, and W.~Wang, ``Logicvista: Multimodal llm logical reasoning benchmark in visual contexts,'' \emph{arXiv preprint arXiv:2407.04973}, 2024.

\bibitem{lu2022learn}
P.~Lu, S.~Mishra, T.~Xia, L.~Qiu, K.-W. Chang, S.-C. Zhu, O.~Tafjord, P.~Clark, and A.~Kalyan, ``Learn to explain: Multimodal reasoning via thought chains for science question answering,'' \emph{Advances in Neural Information Processing Systems (NeurIPS)}, vol.~35, pp. 2507--2521, 2022.

\bibitem{chen2024we}
L.~Chen, J.~Li, X.~Dong, P.~Zhang, Y.~Zang, Z.~Chen, H.~Duan, J.~Wang, Y.~Qiao, D.~Lin \emph{et~al.}, ``Are we on the right way for evaluating large vision-language models?'' \emph{Advances in Neural Information Processing Systems (NeurIPS)}, vol.~37, pp. 27\,056--27\,087, 2024.

\bibitem{yue2024mmmupro}
X.~Yue, T.~Zheng, Y.~Ni, Y.~Wang, K.~Zhang, S.~Tong, Y.~Sun, B.~Yu, G.~Zhang, H.~Sun \emph{et~al.}, ``Mmmu-pro: A more robust multi-discipline multimodal understanding benchmark,'' \emph{arXiv preprint arXiv:2409.02813}, 2024.

\bibitem{cui2025entropy}
G.~Cui, Y.~Zhang, J.~Chen, L.~Yuan, Z.~Wang, Y.~Zuo, H.~Li, Y.~Fan, H.~Chen, W.~Chen \emph{et~al.}, ``The entropy mechanism of reinforcement learning for reasoning language models,'' \emph{arXiv preprint arXiv:2505.22617}, 2025.

\bibitem{fu2025srft}
Y.~Fu, T.~Chen, J.~Chai, X.~Wang, S.~Tu, G.~Yin, W.~Lin, Q.~Zhang, Y.~Zhu, and D.~Zhao, ``Srft: A single-stage method with supervised and reinforcement fine-tuning for reasoning,'' \emph{arXiv preprint arXiv:2506.19767}, 2025.

\bibitem{yu2025dapo}
Q.~Yu, Z.~Zhang, R.~Zhu, Y.~Yuan, X.~Zuo, Y.~Yue, W.~Dai, T.~Fan, G.~Liu, L.~Liu \emph{et~al.}, ``Dapo: An open-source llm reinforcement learning system at scale,'' \emph{arXiv preprint arXiv:2503.14476}, 2025.

\end{thebibliography}

\clearpage

\appendix
\subsection{Prompts}

\label{appendix:a}
\begin{figure}[h]
\centering
\includegraphics[width=.98\linewidth]{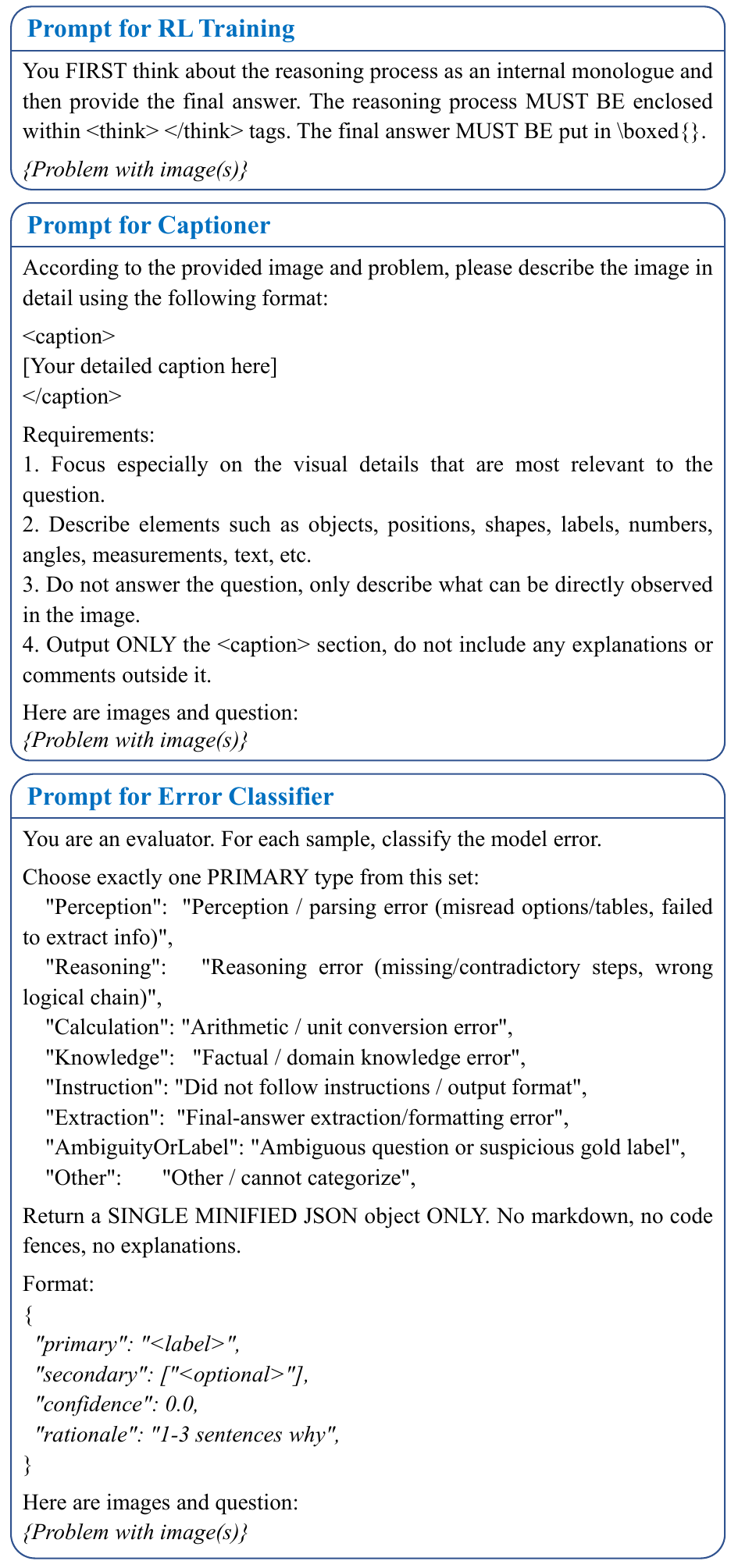}
\caption{Comprehensive overview of the designed prompts employed in our study, covering training, caption generation, and error classification tasks.}
\label{fig_prompt}
\vspace{-1em}
\end{figure}

In Fig.~\ref{fig_prompt}, we summarize all the prompts used in our experiments. 
These prompts provide consistent supervision for training, perception grounding, and error analysis, serving as the basis of our experimental framework. 
It is worth noting that our evaluation does not strictly rely on the given training prompts. Since the benchmark datasets adopt diverse prompt formats, the results remain reliable and demonstrate that CapPO is effective under varying evaluation conditions.

\subsection{Experiment Details}
\label{appendix:b}

\subsubsection{Benchmarks}
The dataset sizes and task focuses are summarized in Table~\ref{tab:benchmarks_overview}, which mainly consists of several hundred to several thousand visual question answering (VQA) and multiple-choice question (MCQ) samples.

For fair comparison, we adopt the publicly released checkpoints of all baseline models and evaluate them using \href{https://github.com/open-compass/VLMEvalKit}{\texttt{VLMEvalKit}} with greedy decoding under \texttt{bf16} precision.
All results are reported in terms of average accuracy under the default settings of \href{https://github.com/open-compass/VLMEvalKit}{\texttt{VLMEvalKit}}. 
Specifically, we use the \textit{test\_mini} split for MathVista, MathVerse, and MathVision, the \textit{CoT} split for WeMath, the \textit{test} split for ScienceQA, the \textit{val} split for MMMU, the \textit{10c\_CoT} for MMLU-Pro. Furthermore, the results on MathVerse are averaged over its five subsets, including V-O, V-D, V-I, T-O, and T-D.  

\begin{table}[h]
\caption{Overview of benchmarks used for evaluation.}
\label{tab:benchmarks_overview}
\centering
\resizebox{\columnwidth}{!}{
\setlength{\tabcolsep}{2.5pt}
\begin{tabular}{l c c l}
\toprule
\textbf{Benchmark} & \textbf{Category} & \textbf{Size} & \textbf{Capability Focus} \\
\midrule
MathVista\cite{lu2023mathvista}   & Math & 1000 & Visual math reasoning, charts and plots \\
MathVerse\cite{zhang2024mathverse}   & Math & 3940    & Diagram-centric math reasoning \\
MathVision\cite{wang2024measuring}  & Math & 304  & Competition-level visual math \\
WeMath\cite{qiao2024we}      & Math & 1740 & Hierarchical visual math reasoning \\
DynaMath\cite{zou2024dynamath}    & Math & 5010  & Robustness under controlled variations \\
\midrule
LogicVista\cite{xiao2024logicvista}  & General & 447   & Visual logical reasoning \\
ScienceQA\cite{lu2022learn}   & General & 2017 & Multimodal science QA \\
MMStar\cite{chen2024we}      & General & 1500 & Vision-indispensable reasoning \\
MMMU\cite{yue2024mmmu}        & General & 1050   & College-level multi-discipline QA \\
MMMU-Pro\cite{yue2024mmmupro}    & General & 1730 & Robust multimodal reasoning \\
\bottomrule
\end{tabular}}
\end{table}

\begin{figure}[h]
\centering
\includegraphics[width=\linewidth]{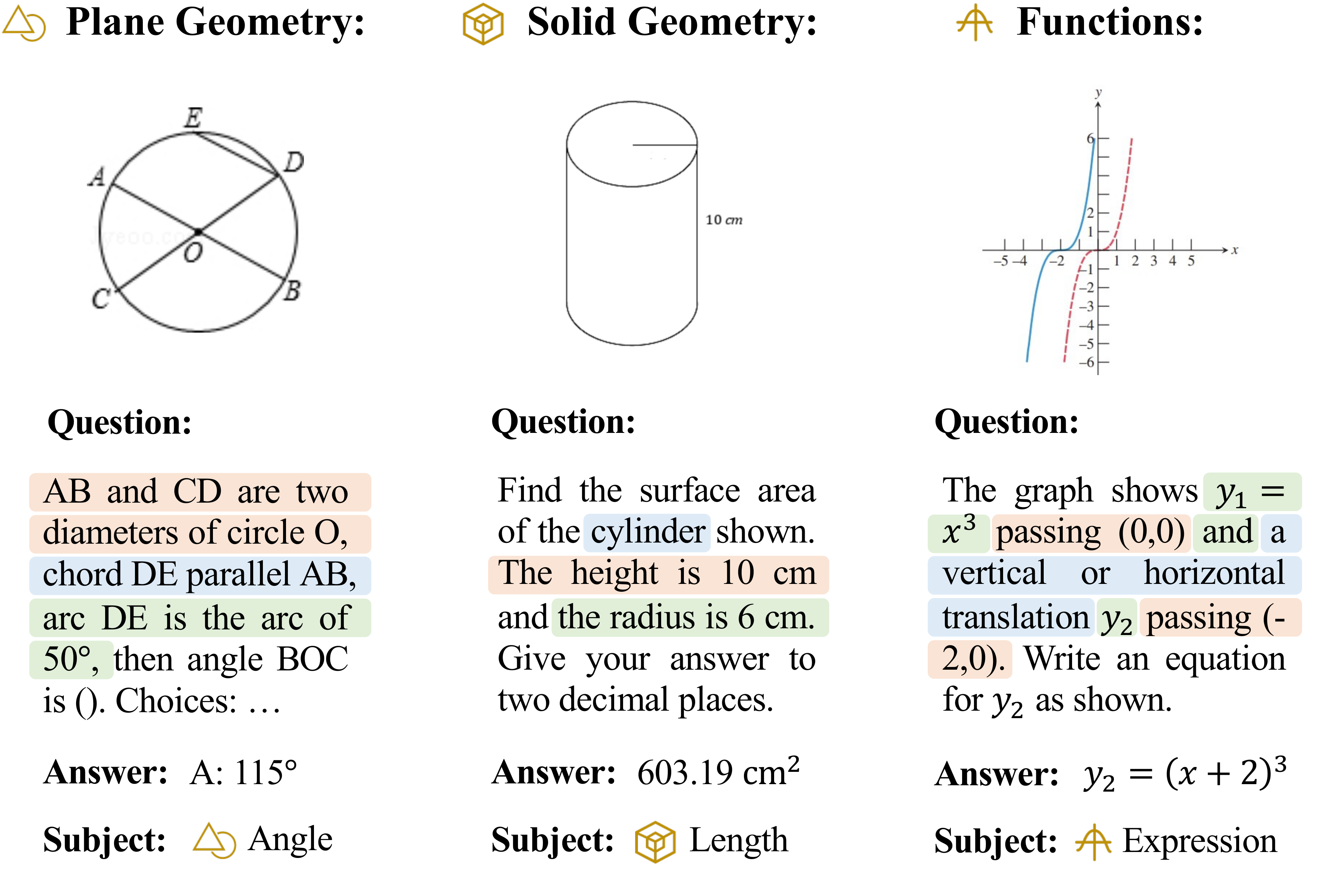}
\caption{Examples of question texts in MathVerse \cite{zhang2024mathverse}.}
\label{fig:mathqa_examples}
\end{figure}

To illustrate, Fig.~\ref{fig:mathqa_examples} shows typical examples from math-related QA datasets (e.g., MathVerse), covering plane geometry, solid geometry, and functional reasoning. These examples highlight the diverse reasoning skills required, including geometric deduction, quantitative calculation, and symbolic expression, thereby reflecting the representative challenges encountered in our evaluation benchmarks.

\subsubsection{Training Details}
We train CapPO using the \href{https://github.com/hiyouga/EasyR1}{\texttt{EasyR1}} framework, 
an efficient and scalable multimodal RL platform designed with a focus on algorithmic flexibility. 
The key training hyperparameters are summarized in Table~\ref{tab:train_params}.
All training scripts and model checkpoints will be released publicly at \href{https://github.com/TU2021/CapPO}{https://github.com/TU2021/CapPO}.

\begin{table}[h]
\caption{Key Training Parameters for CapPO.}
\label{tab:train_params}
\centering
\resizebox{\columnwidth}{!}{
\setlength{\tabcolsep}{4pt}
\begin{tabular}{ll ll}
\toprule
\textbf{Parameter} & \textbf{Value} & \textbf{Parameter} & \textbf{Value} \\
\midrule
Max Prompt Length & 4096 & Max Response Length & 8192 \\
Min Image Pixels & 200,704 & Max Image Pixels & 1,003,520 \\
Rollout Batch Size & 512 & Training Batch Size & 128 \\
Rollout Sampling Num & 8 & Rollout Temperature & 1.0  \\
Clip Coef $\epsilon$ & 0.2 & Learning Rate & 1e-6 \\
Caption-KL Coef $\alpha$ & 0.01 & Advantage-KL Coef $\beta$ & 0.1 \\
Weight Bound $w_{\min}$ & 0.5 & Weight Bound  $ w_{\max}$ & 1.5 \\
Number of Epochs & 2 & Freeze Vision Encoder & False \\
\bottomrule
\end{tabular}
}
\end{table}

\subsection{Computational Cost}
\label{appendix:train_cost}

Most of the computational overhead in RL arises from the rollout stage. 
Relative to GRPO, CapPO introduces only a small number of additional logit computations for caption-based regularization. 
Because CapPO generally produces longer responses, the generation stage exhibits slightly higher latency. 
Nevertheless, the overall training cost remains of the same order of magnitude as GRPO. 
On an 8$\times$H100 cluster, training CapPO requires about 14 hours, compared with roughly 11 hours for GRPO. 
A detailed breakdown of the major per-step time consumption, including response generation, actor update, and each step, is provided in Table~\ref{tab:train_time}.

\begin{table}[h]
\centering
\caption{Training time breakdown of GRPO and CapPO.}
\label{tab:train_time}
\resizebox{1\columnwidth}{!}{
\setlength{\tabcolsep}{2.5pt}
\begin{tabular}{lcccc}
\toprule
\textbf{Method} & \textbf{Generation} & \textbf{Update-Actor} & \textbf{Caption-Logit} & \textbf{Each-Step} \\
\midrule
GRPO   & $\sim$56s  & $\sim$167s &  -  & $\sim$300s  \\
\textbf{CapPO}  & $\sim$73s  & $\sim$183s & $\sim$26s & $\sim$370s  \\
\bottomrule
\end{tabular}
}
\end{table}

\subsection{Discussion: Perception vs. Consistency}

A natural question is why we do not simply adopt stronger vision encoders to reduce recognition errors. While this can improve perception, it often requires costly retraining with massive supervision and cannot fully prevent errors from propagating through reasoning. In contrast, our work targets the RL stage, where perceptual inconsistencies can directly compromise reasoning reliability.

By introducing caption-based consistency regularization, CapPO constrains the policy to align perceptual cues with reasoning outcomes. This brings two key benefits: (1) reduced sensitivity to residual perception errors by penalizing inconsistent trajectories, and (2) efficiency by leveraging captions readily available from MLLMs, avoiding retraining or upgrading the encoder.

Rather than treating perception errors in isolation, our approach addresses the underlying misalignment between perception and reasoning, while remaining complementary to future advances in vision backbones.

\vfill

\end{document}